\documentclass[twocolumn,aps,prl,superscriptaddress]{revtex4}
\usepackage{amsmath,amssymb,mathrsfs,bm}
\usepackage{graphicx,dcolumn,times}
\usepackage[colorlinks,linkcolor=blue,anchorcolor=blue,citecolor=blue,filecolor=blue,menucolor=blue,runcolor=blue,urlcolor=blue,frenchlinks=blue]{hyperref}
\usepackage{amsmath}
\usepackage{graphicx}
\usepackage{setspace}
\usepackage{amssymb}
\usepackage{dcolumn}
\usepackage{mathrsfs}
\usepackage{bm}
\usepackage{lipsum}
\begin{document}

\title{Synthetic Topological Vacua of Yang-Mills Fields in Bose-Einstein Condensates}

\author{Jia-Zhen Li}
\thanks{These authors contributed equally.}
\affiliation{Guangdong Provincial Key Laboratory of Quantum Engineering and Quantum Materials, School of Physics and Telecommunication Engineering, South China Normal University, Guangzhou 510006, China}

\author{Cong-Jun Zou}
\thanks{These authors contributed equally.}
\affiliation{Guangdong Provincial Key Laboratory of Quantum Engineering and Quantum Materials, School of Physics and Telecommunication Engineering, South China Normal University, Guangzhou 510006, China}

\author{Yan-Xiong Du}
\affiliation{Guangdong Provincial Key Laboratory of Quantum Engineering and Quantum Materials, School of Physics and Telecommunication Engineering, South China Normal University, Guangzhou 510006, China}

\author{Qing-Xian Lv}
\affiliation{Guangdong Provincial Key Laboratory of Quantum Engineering and Quantum Materials, School of Physics and Telecommunication Engineering, South China Normal University, Guangzhou 510006, China}

\author{Wei Huang}
\affiliation{Guangdong Provincial Key Laboratory of Quantum Engineering and Quantum Materials, School of Physics and Telecommunication Engineering, South China Normal University, Guangzhou 510006, China}

\author{Zhen-Tao Liang}
\affiliation{Guangdong Provincial Key Laboratory of Quantum Engineering and Quantum Materials, School of Physics and Telecommunication Engineering, South China Normal University, Guangzhou 510006, China}

\author{Dan-Wei Zhang}
\affiliation{Guangdong Provincial Key Laboratory of Quantum Engineering and Quantum Materials, School of Physics and Telecommunication Engineering, South China Normal University, Guangzhou 510006, China}

\affiliation{ Guangdong-Hong Kong Joint Laboratory of Quantum Matter, Frontier Research Institute for Physics, South China Normal University, Guangzhou 510006, China}

\author{Hui Yan}
\affiliation{Guangdong Provincial Key Laboratory of Quantum Engineering and Quantum Materials, School of Physics and Telecommunication Engineering, South China Normal University, Guangzhou 510006, China}
\affiliation{ Guangdong-Hong Kong Joint Laboratory of Quantum Matter, Frontier Research Institute for Physics, South China Normal University, Guangzhou 510006, China}
\affiliation{Guangdong Provincial Engineering Technology Research Center for Quantum Precision Measurement, South China Normal University, Guangzhou 510006, China}

\author{Shanchao Zhang}
\email{sczhang@m.scnu.edu.cn}
\affiliation{Guangdong Provincial Key Laboratory of Quantum Engineering and Quantum Materials, School of Physics and Telecommunication Engineering, South China Normal University, Guangzhou 510006, China}
\affiliation{ Guangdong-Hong Kong Joint Laboratory of Quantum Matter, Frontier Research Institute for Physics, South China Normal University, Guangzhou 510006, China}

\author{Shi-Liang Zhu}
\email{slzhu@scnu.edu.cn}
\affiliation{Guangdong Provincial Key Laboratory of Quantum Engineering and Quantum Materials, School of Physics and Telecommunication Engineering, South China Normal University, Guangzhou 510006, China}
\affiliation{ Guangdong-Hong Kong Joint Laboratory of Quantum Matter, Frontier Research Institute for Physics, South China Normal University, Guangzhou 510006, China}


\begin{abstract}
Topological vacua are a family of degenerate ground states of Yang-Mills fields with zero field strength but nontrivial topological structures. They play a fundamental role in particle physics and quantum field theory, but have not yet been experimentally observed. Here we report the first theoretical proposal and experimental realization of synthetic topological vacua with a cloud of atomic Bose-Einstein condensates. Our setup provides a promising platform to demonstrate the fundamental concept that a vacuum, rather than being empty, has rich spatial structures. The Hamiltonian for the vacuum of topological number $n=1$ is synthesized and the related Hopf index is measured. The vacuum of topological number $n=2$ is also realized, and we find that  vacua with different topological numbers have distinctive spin textures and Hopf links. Our work opens up opportunities for exploring topological vacua and related long-sought-after instantons in tabletop experiments.

\end{abstract}
\maketitle

\emph{Introduction.}---The vacuum of a gauge field is the field state with the lowest energy and thus zero field strength. It is crucial to our understanding of some amazing features of particle structures and quantum fields~\cite{Weinberg1996,Bick-Steffen}. For example, in quantum electrodynamics, a familiar vacuum is the zero-point fluctuation of the electromagnetic field, which leads to important physical effects such as the Lamb shift, anomalous magnetic moment of electrons, and Casimir force. As another example,  in the electroweak theory, the investigation of vacuum leads to our deep understanding of the spontaneous symmetry breaking of vacuum, the origin of mass, and the predication of the Higgs bosons. Furthermore, the studies of the quantum chromodynamics vacuum  may explain the
origin and value of the quark and gluon condensates, the mechanism
for quark confinement and chiral symmetry breaking~\cite{Weinberg1996,Bick-Steffen}.

 The non-Abelian Yang-Mills field is predicted to have an intriguing type of topological vacua (TV), namely, a family of degenerate ground states   with zero field strength but nontrivial topological structures~ \cite{Belavin1975,Hooft1976,Jackiw1976,Bick-Steffen}. This type of vacua is predicted to be significantly different from an Abelian vacuum in terms of energy degeneracy and field topology. Furthermore, quasiparticles called instantons  are known to describe tunneling processes between different vacua, which lead to the introduction of the $CP$-violating $\theta$ term and shed light on the mechanism of quark confinement \cite{Weinberg1996,Bick-Steffen}. These nonperturbative solutions of the Yang-Mills fields were first proposed in 1975~\cite{Belavin1975}, but have not yet been experimentally observed.

Inspired by the realization of the Abelian Higgs model with Bose-Einstein condensates (BECs)~\cite{Leonard2017,Endres2012} and the rapidly developing quantum simulation of synthetic gauge fields in ultracold atoms \cite{YJLin2009,Beeler2013,Duca2015,ZWu2016,LHuang2016,Kolkowitz2016,Sugawa2018,Fletcher2021,QXLv2021,Dalibard2011,DWZhang2018} and other quantum-engineered systems~\cite{XTan2021,MChen2022}, in this work, we theoretically propose and experimentally realize the first scheme to synthesize TV of Yang-Mills fields using a cloud of ultracold atomic BECs coupled with a pair of Raman laser fields. The exemplary TV with topological number (TN) $n=1$ is realized and the related Hopf index is measured. Furthermore, both the three-dimensional spin textures and Hopf links of this family of TV with $n=1$ and $2$ are demonstrated. Our work opens up a potential way to study TV  with engineered quantum systems.

\begin{figure}[ptb]
\begin{center}
\includegraphics[width=7.5cm]{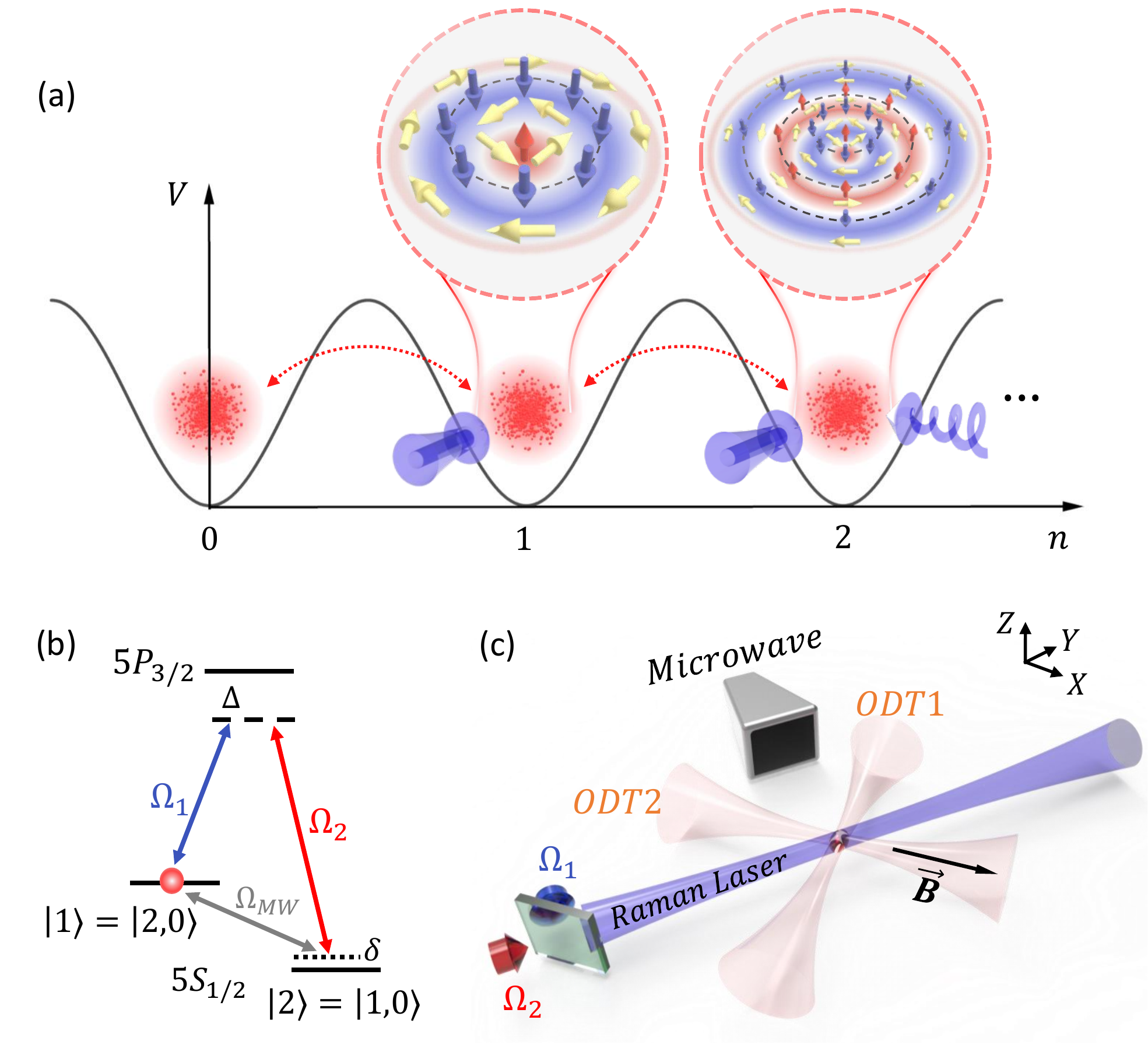}
\caption{\label{fig:model}
Scheme of a family of TV synthesized with ultracold atoms and  experimental realization.
(a) Schematic plot of the potential energy $V[\mathbf{A}]$ as a function of the TN $n$. The vacuum with fixed TN has specific spin textures. Quantum tunneling between vacua is allowed with the assistance of instantons.
(b) Proposed atomic energy level configuration. Two Zeeman sublevels that belong to the ground hyperfine manifolds of $^{87}$Rb atoms are selected: $|1\rangle=|5~^1S_{1/2},F=2,m_F=0\rangle$ and $|2\rangle=|5~^1S_{1/2},F=1,m_F=0\rangle$, which are cyclically coupled via a detuned two-photon Raman process ($\Omega_{1,2}$) together with a microwave ($\rm \Omega_{MW}$).
(c) Experiment setup. A cloud of $\rm ^{87}Rb$ BEC are trapped in a crossed optical dipole trap (ODT). A homogeneous magnetic field along the $x$ axis sets the quantum axis. A pair of copropagating Raman lasers $\Omega_{1,2}$ focused at the cloud couple the two states $|1\rangle$ and $|2\rangle$. A microwave field emitted from a microwave waveguide is used to realize  the initial state preparation and synthesized vacua measurement.}
\end{center}
\end{figure}

\emph{Topological vacua of the Yang-Mills fields.}---A gauge field is fully described by a gauge potential $ {A}_{\mu}$. The commonly studied kinetic effects of a field are determined by the field strength  $ {F}_{\mu\nu}\equiv \partial_{\mu} {A}_{\nu}-\partial_{\nu} {A}_{\mu}-i [{A}_{\mu},  {A}_{\nu}]$. Notably, a gauge field  also has geometric effects, such as the Aharonov-Bohm effect, which are solely determined by $ {A}_{\mu}$. It is worth noting that for the same $ {F}_{\mu\nu}$,  there exist a group of gauge potentials $ {A}_{\mu}$'s  that are connected by gauge transformations described by $N\times N$ unitary matrices $ {U}(N)$. With an Abelian vacuum, ${F}_{\mu\nu}=0$ results in the only choice of $ {A}_{\mu}=0$. However, for a non-Abelian $ {U}(N)$, besides the regular vacuum with $ {A}_{\mu}=0$, there are a group of $ {A}_{\mu}= {U}^{-1}\partial_{\mu} {U}$ gauge potentials that possess rich geometric effects.  Searching for nonperturbative solutions of the SU(2) Yang-Mills fields reveals a family of solutions described by  the gauge transformations $ {U}_n=( {U}_1)^n$ with $n$ being  an integer~\cite{Jackiw1976,Hooft1976,Belavin1975}, where
\begin{equation}\label{U1Mat}
 {U}_1(\mathbf{r})=\frac{r^2-\eta^2}{r^2+\eta^2} {\sigma_0}-\frac{2i\eta\mathbf{r}\cdot {\vec{\sigma}}}{r^2+\eta^2}.
\end{equation}
Here $\vec{\sigma}=(\sigma_x,\sigma_y,\sigma_z)$ are Pauli matrices, $ {\sigma_0}$ is the identity matrix, $\mathbf{r}=(x,y,z)$ is the position vector with $r=|\mathbf{r}|$, and $\eta$ is a constant. As can be seen,  the field strength $ {F}_{\mu\nu}$  related to gauge transformation $ {U}_n=( {U}_1)^n$  vanishes and thus they represent degenerate vacua. Interestingly, these vacua are  characterized by a TN,
\begin{equation}\label{Windingnumber}
n=\frac{1}{24\pi^2}\int_{-\infty}^{+\infty} d\mathbf{r}\epsilon_{ijk} { \rm{Tr}} ( {U}^{-1}_n\partial_{i} {U}_n {U}^{-1}_n\partial_{j}  {U}_n {U}^{-1}_n\partial_{k} {U}_n),
\end{equation}
where $\epsilon_{ijk}$ is the Levi-Civita symbol. 
As shown in Fig.~\ref{fig:model}(a), there must exist potential barriers to separate vacua with different TNs. Accompanied by instantons, quantum transition between vacua is predicted\cite{Hooft1976,Weinberg1996,Bick-Steffen}.

\emph{ Artificial gauge fields for a light-atom system.}---Here we use atomic BECs to demonstrate that TV can be synthesized with an engineered  quantum system. The Hamiltonian of a laser-atom interaction system reads $H=\frac{\mathbf{p}^{2}}{2m}+V(\mathbf{r})+H_{\text{AL}}$, where $m$ is the atomic mass, the laser-atom interaction $H_{\text{AL}}$  is an $N\times N$
matrix in the basis of the internal energy levels $|j\rangle$, and the potential $V(\mathbf{r})=\sum_{j=1}^N V_j(\mathbf{r})|j\rangle \langle j|$. In this case, the full quantum state can  be expanded as $|\Phi (\mathbf{r})\rangle =\sum_{j=1}^{N}\phi_{j}(\mathbf{r})|j\rangle $.

In the representation of the dressed states $|\chi_n\rangle$ that are eigenvectors of the Hamiltonian
$H_{\text{AL}}$, $H_{\text{AL}}|\chi_n\rangle=\varepsilon_n|\chi_n\rangle$, the full quantum state of the atom $|\tilde{\Phi} (\mathbf{r})\rangle $ is written as $|\tilde{\Phi} (\mathbf{r})\rangle =\sum_{j}\psi_{j}(\mathbf{r})|\chi _{j}(\mathbf{r})\rangle $, where the wave functions $|\tilde{\Psi}\rangle =(|\psi _{1}\rangle,|\psi _{2}\rangle,\ldots,|\psi_{N}\rangle)^{\top}$ obey the Schr\"{o}dinger equation $i\hbar\frac{\partial}{\partial {t}}|\tilde{\Psi}\rangle =\tilde{H}_{\text{eff}}|\tilde{\Psi}\rangle $, with the effective Hamiltonian $\tilde{H}_{\text{eff}}=UHU^{\dagger}$. We further assume that the first $2$ atomic dressed states among the total $N$ states are degenerate and are well separated from the remaining $N-2$ states. This way, we can project the full Hamiltonian onto this subspace. Under this condition, the wave function in the subspace
$\Psi=\left(\psi_1,\psi_2\right)^\top$ is again governed by the Schr\"odinger equation $i\hbar\frac{\partial}{\partial t}\Psi={H}_{\text{eff}}\Psi$ with the effective Hamiltonian  taking the following form~\cite{Berry1984,Wilczek1983,CPSun1990,Ruseckas2005,SLZhu2006}:
\begin{equation}
H_{\text{eff}}=\frac{1}{2m}
(-i\hbar\nabla -\mathbf{A})^2 + {V}_\text{eff}. \label{H_non}
\end{equation}
Here $\mathbf{A}=i\hbar U\nabla U^{\dagger }$ with $U=(|\chi_1 \rangle,|\chi_2)\rangle)$ and ${V}_\text{eff}$ is a scalar potential in Supplemental Material (SM)~\cite{SM}.

\emph{ Realizing topological vacua with atoms.}---Topological vacua can be realized using ultracold atoms with two fully (or almost) degenerate states  by designing the laser-atom interactions. Two degenerate states can be achieved with four-level atoms, such as a tripod-level or $\infty$-level  configuration~\cite{DWZhang2018,QXLv2021}. For experimental simplicity, we propose a feasible scheme with two almost degenerate states. We consider three-level atoms cyclically coupled by three position-dependent fields $\Omega_{1,2}(\mathbf{r})$ and $\rm \Omega_{MW}(\mathbf{r})$, as shown in Fig.~\ref{fig:model}(b). The Hamiltonian $ {H}_{\text{AL}}(\mathbf{r})$ can be written as
\begin{equation}\label{H_int}
 {H}_{\text{AL}}(\mathbf{r})=\frac{\hbar}{2}
\left(\begin{array}{cccc}
0& \Omega^*_{MW}(\mathbf{r}) & \Omega^*_1(\mathbf{r}) &\\
\Omega_{MW}(\mathbf{r}) & -2\alpha(\mathbf{r})|\delta(\mathbf{r})| & \Omega^*_2(\mathbf{r}) &\\
\Omega_1(\mathbf{r}) & \Omega_2(\mathbf{r}) & 2\alpha(\mathbf{r})|\Delta(\mathbf{r})| &
\end{array}\right),
\end{equation}
where $\Delta(\mathbf{r})$ [$\delta(\mathbf{r})$] is the single-photon [two-photon] detuning and $\alpha(\mathbf{r})={\rm sgn}(|\mathbf{r}|-1)$ is a sign function.

In order to easily obtain a solution for the TV, we assume
$\Omega_{\rm MW}(\mathbf{r})=-i\Omega_1(\mathbf{r})\sqrt{|\delta(\mathbf{r})/\Delta(\mathbf{r})|}.$
To simplify the notations, we  hide the notation $\mathbf{r}$ later on. We solve the Schr\"odinger equation $H_{\text{AL}}|\chi_n\rangle=\varepsilon_n|\chi_n\rangle$ under the large detuning condition, i.e., $\Delta \gg |\delta|,\Omega,$ with $\Omega=\sqrt{|\Omega_1|^2+|\Omega_2|^2},$  and then obtain the eigenvalues $\varepsilon \approx 0,-\hbar\kappa^2/(4\Delta), \hbar[\Delta+\Omega^2/(4\Delta)]$ with $\kappa=\sqrt{\Omega^2+4|\delta\Delta|}$. The first two are nearly degenerate since there is a large gap between the first two  and the last one. They may create a subspace for synthetic SU(2) gauge field. To clearly present that $\{|\chi_1\rangle,|\chi_2\rangle\}$ form a pseudospin subspace, we denote them as $\{|\chi_{-}\rangle,|\chi_{+}\rangle\}$.   After solving the related eigenvectors, we derive  the transformation
\begin{equation}\label{U_int}
U_{\text{AL}}=\frac{1}{\kappa}
\left(\begin{array}{cc}
-i\Omega_2+2\alpha\sqrt{|\delta\Delta|} & i\Omega_1^* \\
i\Omega_1 & i\Omega_2+2\alpha\sqrt{|\delta\Delta|}
\end{array}\right).
\end{equation}
 Therefore, if we can find a solution with $U_{\text{AL}}=U_n$, then a TV with  TN $n$ is realizable. In SM~\cite{SM}, we show such solutions. In particular, we find that $\Omega=2\sqrt{|\delta\Delta|}\tan[2\arctan(r/\eta)]$ leads to $U_{\text{AL}}=U_{1}$. We can further realize vacua with different TN $n$ in an array configuration as shown in Fig. \ref{fig:model}(a), and then instantons can emerge in such an array~\cite{Note1}.

 We can realize the SU(2) Yang-Mills vacua with desired $ {U}_n(\mathbf{r})$ if we implement the laser-atom interaction to obtain the topologically equivalent  Hamiltonian $H_\text{TV} (\mathbf{r})=U_n(\mathbf{r})\sigma_z U_n^{\dagger}(\mathbf{r})$. However, engineering such a Hamiltonian in real space is challenging since the strength, frequency, and phase of the coupling fields in each spatial point must be well designed.  In the first experiment, we simply implement the above $ {H}_\text{TV}(\mathbf{r})$ in a parameter space and then measure the TN and the related significant properties of the TV.
Our experiment can visualize the spatial structure of the TV, which has not been explored in previous literature.

The TN in Eq.(\ref{Windingnumber}) is equivalent to the Hopf index $\chi_{\rm Hopf}$\cite{Wilczek1983,Moore2008,SM,XXYuan2017},
\begin{equation}\label{hopfidx}
n=\chi_{\rm Hopf}=-\frac{1}{4\pi^2}\int_{-\infty}^{+\infty} d\mathbf{r}{\mathbf{a}}(\mathbf{r})\cdot{\mathbf{f}}(\mathbf{r})
\end{equation}
where the $k$th component of $\mathbf{a}(\mathbf{r})$ defined as $\mathbf{a}_k(\mathbf{r})=\langle{\chi_-(\mathbf{r})}\partial_{k}|{\chi_-(\mathbf{r})}\rangle$ is  Berry connection
and $\mathbf{f}(\mathbf{r})=\nabla \times \mathbf{a}(\mathbf{r})$ is Berry curvature \cite{Berry1984}.
Experimentally, we can obtain both $\mathbf{a}(\mathbf{r})$ and $\mathbf{f}(\mathbf{r})$ by adiabatically tuning the Raman laser fields and detecting the  state $|{\chi_-(\mathbf{r})}\rangle$. According to Eq.(\ref{hopfidx}), the TN can be measured by detecting the spin states of the atoms in the full space of $\mathbf{r}$. Furthermore, the properties of the vacuum can be determined by the density matrices  $\langle {\mathbf{\sigma}}\rangle=\{
\langle{\chi_-}| {\sigma}_x|{\chi_-}\rangle,
\langle{\chi_-}| {\sigma}_y|{\chi_-}\rangle,
\langle{\chi_-} |{\sigma}_z|{\chi_-}\rangle\}$~\cite{SM}.

\begin{figure*}[ptb]
\begin{center}
\includegraphics[width=13.0cm]{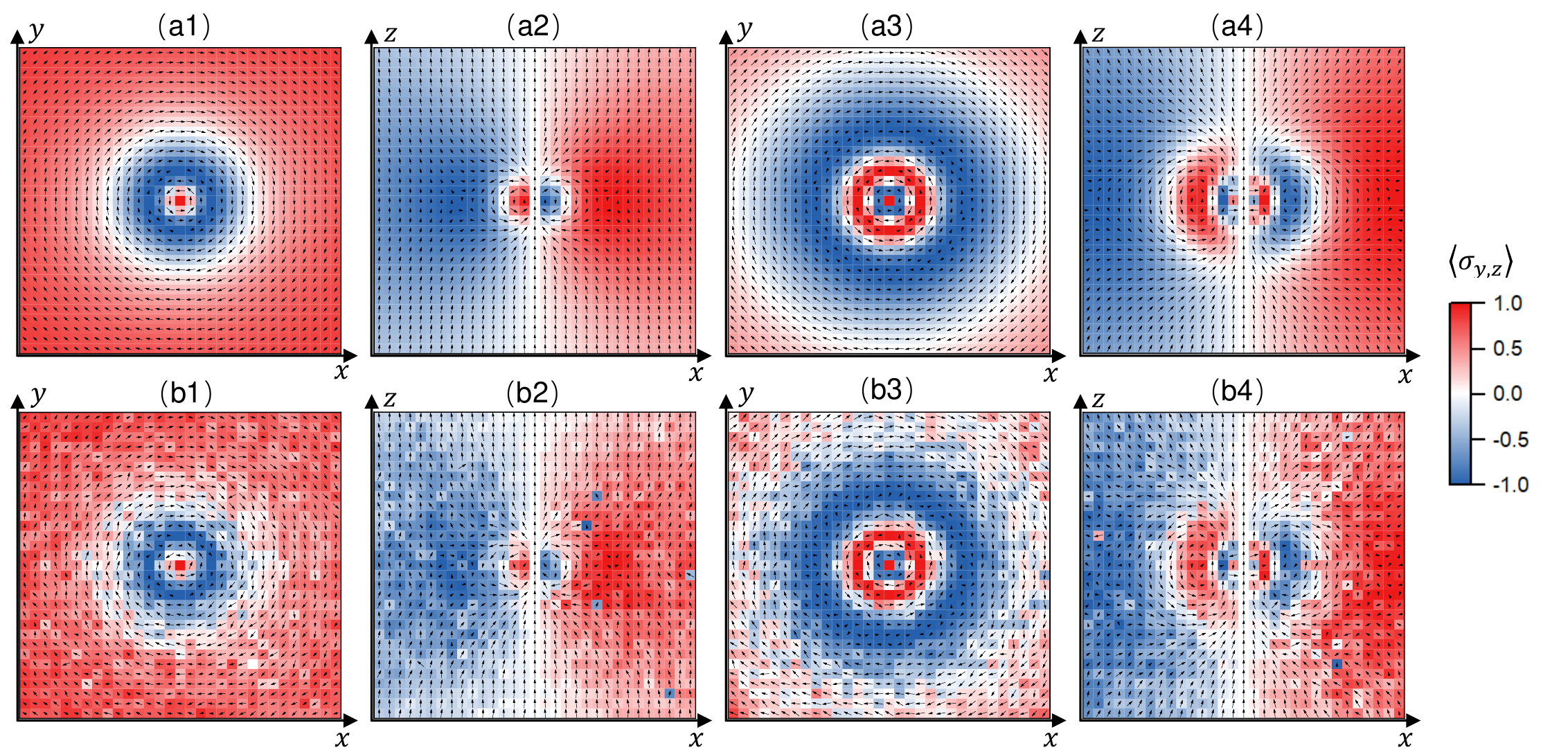}
\caption{\label{fig:spin}  The topologies of vacua visualized by spatial distribution of atomic spin direction.
(a1)$-$(a4) Theoretical results of density matrices that show the direction of atomic pseudospin.  (b1)$-$(b4) Experimentally measured results.
Panels (a1) and (b1) correspond to the spin texture of the vacuum with $n=1$ in the plane of $z=0$, while panels (a2) and (b2) are those in the plane of $y=0$.
Panels (a3) and (b3) correspond to the spin texture of the vacuum with $n=2$ in the plane of $z=0$, while panels (a4) and (b4) are those in the plane of $y=0$.
Black arrows represent the in-plane components, while colors (with  values mapped to the right-hand-side color bar) indicate the remaining components perpendicular to the plane.
}
\end{center}
\end{figure*}

\emph{Experimental scheme.}---The schematics of our experiment setup is shown in Fig.~\ref{fig:model}. A cloud of $\rm ^{87}Rb$ atoms are laser cooled in a magneto-optical trap and then  evaporatively cooled down to  BEC state in a far off-resonant crossed optical dipole trap (ODT). A weak homogeneous magnetic field $\bold{B}$ along the $x$ axis sets a quantum axis. In order to maintain a long coherence time, we choose two quantum states $|{1}\rangle$ and $|{2}\rangle$ from the magnetic insensitive hyperfine Zeeman sublevels  to mimic the pseudospin.  Initially, all atoms are polarized in pseudospin state $|1\rangle$ by a coherent microwave pulse.

The expected  Hamiltonian $H_\text{TV}(\mathbf{r})$ at specified position $\mathbf{r}$ is realized by adding two Raman lasers with respective Rabi frequency $\Omega_1$ and $\Omega_2$. As shown in the energy level configuration in Fig.~\ref{fig:model}(b), the paired Raman lasers couple the two spin states  $|{1}\rangle$ and $|{2}\rangle$ via two-photon process. The effects of excited states  are adiabatically eliminated by setting  $\Delta=2\pi\times 3.9$ THz since  both  $\Omega_{1,2}$ and  $\delta$ are on the order of $2\pi\times 10$ kHz. The third coupling field $\Omega_{\rm MW}$ is estimated to be around $1$ Hz and hence can be safely omitted. Therefore,  the two quantum states  $|1\rangle$ and $|2\rangle$, together with the paired Raman lasers, produce an approximate degenerate sub-Hilbert space.

To detect the state $|\chi_{-}\rangle$ at each position $\mathbf{r}$, we manipulate the Hamiltonian $ H_\text{TV}$ adiabatically and  drive the atoms from the initial state $|1\rangle$ to the final state $|\chi_{-}\rangle$ in an adiabatic way. In our experiment, the largest Rabi frequency $\Omega_M$ is $2\pi\times 28.5$ kHz, while the coherence time of spin states is longer than $8$ ms. For adiabatic state evolution, the single photon detuning $\Delta$ is kept almost constant while the Raman coupling strength $\Omega_R$ and two-photon detuning $\delta$ ramp smoothly from the respective initial value of $\Omega_R=0$ and $\delta=\Omega_M$. The whole evolution time is set to  around $350$ $\mu s$ ( $\sim 20\pi/\Omega_M$), which is longer compared to the typical time of Rabi oscillation but much shorter than the coherence time and thus ensures the adiabatic and coherent state evolution. With this method, we may adiabatically prepare one Hamiltonian
$H_\text{TV}(\mathbf{r})$ at arbitrary parameter $\mathbf{r}$ in a single experiment run, during which atoms are loaded into the expected state $|\chi_{-}\rangle$   and ready for detections.

\begin{figure}[ptb]
\begin{center}
\includegraphics[width=7.0cm]{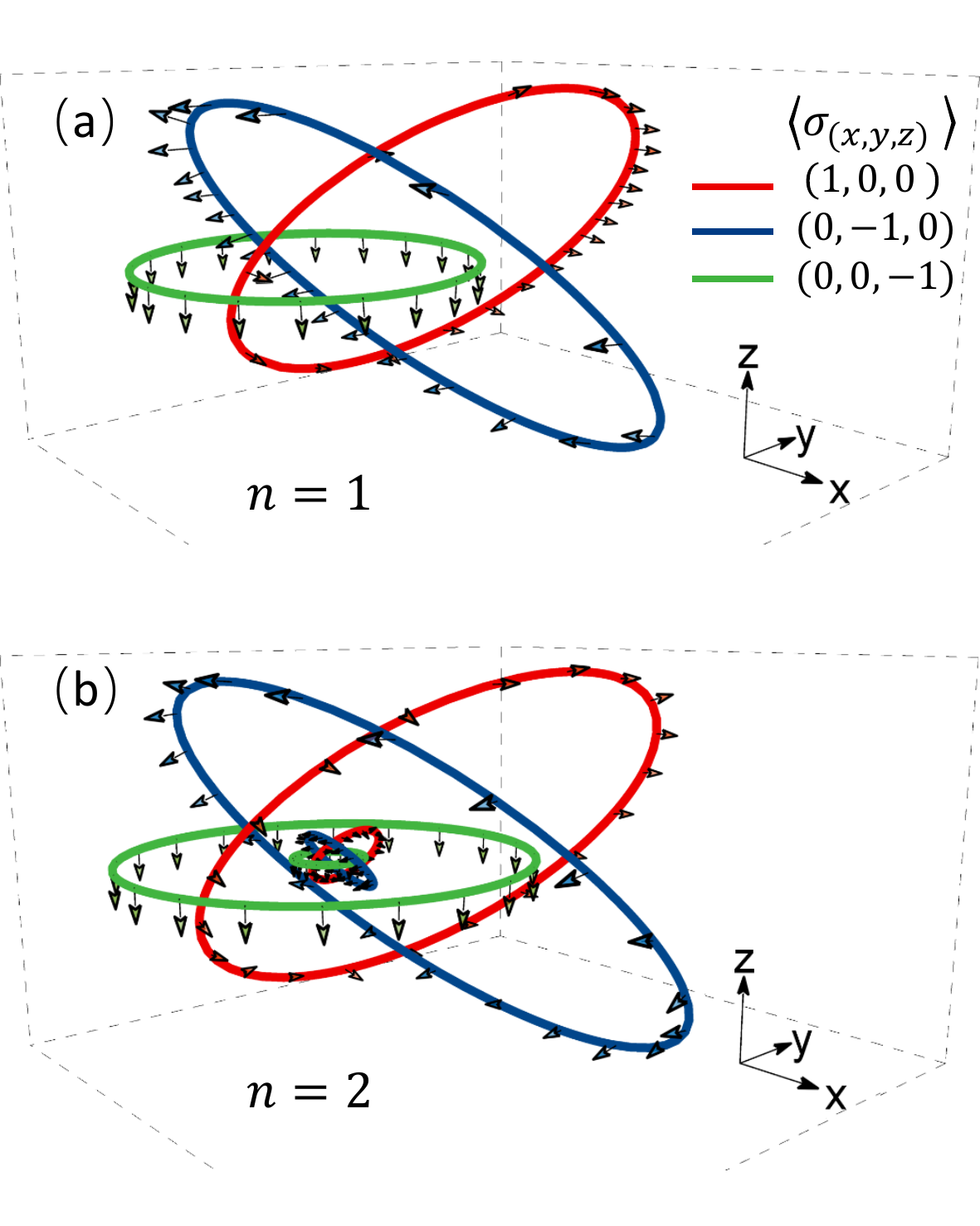}
\caption{\label{fig:link} Hopf links of topological vacua. Hopf links with (a) $n=1$ and (b) $n=2$.
Hopf links belonging to spin directions in the positive $x$, negtive $y$, and negtive $z$ axis are plotted in red, blue, and green, respectively. Solid lines are theoretical results, and arrows denote the spin direction determined by the measured data.}
\end{center}
\end{figure}

\emph{ Measuring the topological number.}---We first synthesize the TV of $n=1$. Along all three ${x,y,z}$ directions, we take the step size $0.1$ and range of $\mathbf{r}$ as  ${-3\leq x,y,z\leq3}$. Here and in the following, the nonzero parameter $\eta$ in Eq.(1) is taken as the length unit of $x,y,z$~\cite{Note2}. To detect TN,  we achieve the Berry curvature $\mathbf{f}(\mathbf{r})$ point by point by measuring the density matrix of atoms at each point in the parameter space. Because of the gauge choice problem, it is difficult to directly measure the Berry connection $\mathbf{a}(\mathbf{r})$. As long as a specified gauge is chosen, the Berry connection at each point could be derived from the measured distribution of Berry curvature $\mathbf{f}(\mathbf{r})$. Eventually, the TN can be obtained by summing up the inner product of $\mathbf{f}(\mathbf{r})$ and $\mathbf{a}(\mathbf{r})$ at all the measured points~\cite{SM}. By repeating the Hamiltonian $ {H}_\text{TV}$ preparation and spin density matrix measurement at each point of $\mathbf{r}$, we measure the Hopf index of $ {U}_1$  as $n=0.91$, which is limited by the average Hamiltonian preparation fidelity of $0.97\pm0.03$. The fidelity is evaluated according to the measured density matrices at all points, and the error is standard deviation.

\emph{Spin texture.}---Intriguingly, our setup provides a unique platform to demonstrate that vacua have rich spatial structures. We reveal that the distribution of $ {U}(\mathbf{r})$ can be used to visualize the topological structure of a  vacuum and that the vacua with different TNs have distinctive spatial spin textures. As $ {U}(\mathbf{r})=(|{\chi_+(\mathbf{r})}\rangle,|{\chi_-(\mathbf{r})}\rangle)$ in our experiment and $|{\chi_+(\mathbf{r})}\rangle,|{\chi_-(\mathbf{r})}\rangle$ are always orthogonal to each other, the spin state $|{\chi_-(\mathbf{r})}\rangle$ can well demonstrate the properties of  the synthesized vacuum. In order to show these structures, we utilize  $\langle {\mathbf{\sigma}}\rangle$  to depict the spatial textures of atomic spins.

  Spin texture in the horizontal $xy$ plane of $z=0$ and vertical $yz$ plane of $x=0$ are measured. The measurements are conducted at each point with grid spacing $0.3$  and the range  $-5\leq x,y,z \leq 5$.  The spin textures of the vacua with $n=1$ and $n=2$ are plotted in Figs. \ref{fig:spin}(a1)$-$\ref{fig:spin}(a4). As an example, Fig. \ref{fig:spin} (a1) shows spin textures in the $xy$ plane with $z=0$. The black arrows indicate the direction of the in-plane $xy$ components $\langle {\sigma}_x\rangle$ and $\langle {\sigma}_y\rangle$, while colors depict  the magnitude of $z$ component $\langle {\sigma}_z\rangle$.

The topologies  of the vacua can be intuitively  understood  by checking the rotation of the spin texture. For $n=1$, as shown in Figs. \ref{fig:spin}(a1) and \ref{fig:spin}(b1), all directions of ${\langle {\sigma}_x\rangle,\langle {\sigma}_y\rangle}$, and $\langle {\sigma}_z\rangle$ reverse  one time in space, which means that the spin texture reverses its direction once from the center of the space  to the outside. At both the center and the infinity far away region in Figs. \ref{fig:spin}(a1) and \ref{fig:spin}(b1),  all spins point to the positive $z$ axis with $\langle {\sigma}_z\rangle=1$. In between, there exists a (deep blue) ring shape where all  spins point to the negative $z$ axis with $\langle {\sigma}_z\rangle=-1$. This spin texture suggests that the gauge potential twists one time in space, which can be more clearly seen in the vertical $yz$ plane with $x=0$ from the arrow direction shown in Figs. \ref{fig:spin}(a2) and \ref{fig:spin}(b2).

Spin textures have significant differences between  vacua  $n=1$ and $n=2$, as shown in Figs. \ref{fig:spin}(a3) and \ref{fig:spin}(b3) and in Figs. \ref{fig:spin}(a4) and \ref{fig:spin}(b4). In both cases, the spins  point to the positive $z$ axis both at the center and the infinity far away region, but there are two ring shapes for vacuum $n=2$ where all spins point to the negative $z$ axis. Therefore, the spins twist twice in space for the $n=2$  vacuum~\cite{Note3}.

\emph{ Hopf links.}---We further show that Hopf links are another powerful way to reveal the intrinsic spatial structure of the TV. A Hopf link is a trace of points where the spin points to the same direction. It has been used to investigate topological  solitons \cite{Wilczek1983,Faddeev1999,Tai2018} and the topological properties in the Brillouin zone (a three-torus $\mathcal{T}^3$)~\cite{XXYuan2017,Belopolski2022}. The link space here is an ordinary infinite space $\mathcal{R}^3$. To understand the vacuum structures, we plot exemplary Hopf links for $n=1$ and $n=2$  in the three-dimensional parameter space, as shown in Fig.\ref{fig:link}.  We select three typical directions (positive $x$ axis, negative $y$ axis, and $z$ axis) to plot Hopf links, which are theoretically depicted by the  solid red, blue, and green lines, respectively. The spin texture at a series of points on each line are experimentally measured and shown by arrows in Fig.\ref{fig:link}. For the $n=1$  vacuum, there is only one link for each spin direction and these three Hopf links interwind with each other one time in space.
When the TN of the vacuum is $n=2$, there exist two separate links for each spin direction. The 6 total links are in 2 different groups;  in each group, the links of different spin directions interwind with each other one time. Therefore, the total TNs are equally  contributed by the two groups of spin winding.

\emph{Conclusion.}---In summary, we have reported the first experiment to realize synthetic TV and explored their nontrivial properties.  These results establish the first experimental platform to explore the fundamental structure of TV.  Our work can be extended to other quantum-engineered systems, such as superconducting qubits and trapped ions. Our theoretical scheme can be applied to realizing a three-dimensional real space TV and an array of TV as shown in Fig. ~\ref{fig:model}(a) and in Fig. S1 in  SM~\cite{SM}. Although such experiments are challenging, once they are realized, the  long-sought-after instantons  and nonperturbative features of the Yang-Mills fields can be explored in tabletop experiments. However, just as many work on this direction \cite{Dalibard2011,DWZhang2018,YJLin2009,Beeler2013,Duca2015,ZWu2016,LHuang2016,Kolkowitz2016,Sugawa2018,Fletcher2021,QXLv2021,XTan2021,MChen2022}, our simulated SU(2) gauge field is a kind of fixed classical gauge field felt by particles and it does not have its own dynamics. Combined with the recently developed technologies of creating synthetic gauge fields with its own dynamics \cite{Banuls2020}, our work may shed light on simulating the vacua of the quantized Yang-Mills fields, which is an open question in quantum field theory.

\begin{acknowledgments}
We thank  Y. Q. Zhu for his contributions on the detection of the Hopf index, and D. L. Deng, L. M. Duan, and Y. X. Zhao for helpful discussions. This work was supported by the Key-Area Research and Development Program of GuangDong Province (Grant No. 2019B030330001),   the National Key Research and Development Program of China (Grant No. 2020YFA0309500), and the National Natural Science Foundation of China (Grants  No. U20A2074, No. 12074132,  No. 12074180, No. 12174126, No. 12104168, and No. U1801661).

\end{acknowledgments}

\newpage
\begin{appendix}
\begin{widetext}
\center
\textbf{\large Supplemental Materials:\\Synthetic Topological Vacua of Yang-Mills Fields in Bose-Einstein Condensates}
\end{widetext}

\section{Theoretical scheme}

\subsection{Artificial gauge fields for a light-atom system}

An artificial gauge field can emerge in cold atom systems when the atomic center-of-mass motion is coupled to its internal degrees of freedom through laser-atom interaction \cite{DWZhang2018}.To understand this artificial gauge field, we consider an  adiabatic motion of neutral atoms with $N$ internal levels in laser fields. The full Hamiltonian of the atoms reads
\begin{equation}\label{H_original}
H=\frac{\mathbf{p}^{2}}{2m}+V(\mathbf{r})+H_{\text{AL}},
\end{equation}
where 
the laser-atom interaction $H_{\text{AL}}$  depends on the position of the atoms and is an $N\times N$ matrix in the basis of the internal energy levels $|j\rangle$. In addition, the potential $V(\mathbf{r})$ is assumed to be diagonal in the internal states $|j\rangle$ with the form $V(\mathbf{r})=\sum_{j=1}^N V_j(\mathbf{r})|j\rangle \langle j|$. In this case, the full quantum state of the atoms  can then be expanded to $|\Phi (\mathbf{r})\rangle =\sum_{j=1}^{N}\phi_{j}(\mathbf{r})|j\rangle $.

We may discuss the problem in the representation of the dressed states $|\chi_n\rangle$ that are eigenvectors of the Hamiltonian $H_{\text{AL}}$, that is, $H_{\text{AL}}$ $|\chi_n\rangle=\varepsilon_n|\chi_n\rangle$. Then the dressed states $|\chi \rangle =(|\chi_1 \rangle,|\chi_2 \rangle,\cdots,|\chi_N \rangle)^{\top} $ (with $\top$ denoting the transposition) are related to the original internal states $|j\rangle$ with the relation $|\chi \rangle =U(|1\rangle,|2\rangle ,\cdots,|N\rangle )^{\top}$, where the transform matrix
\begin{equation}\label{Ua}
U=(|\chi_1 \rangle,|\chi_2\rangle,\cdots,|\chi_N \rangle)
\end{equation}
 is an $N\times N$ unitary operator. In the new basis $|\chi \rangle $, the full quantum state of the atom $|\Phi (\mathbf{r})\rangle $ is written as $|\Phi (\mathbf{r})\rangle =\sum_{j}\psi_{j}(\mathbf{r})|\chi _{j}(\mathbf{r})\rangle $, where the wave functions $|\Psi\rangle =(|\psi _{1}\rangle,|\psi _{2}\rangle,\cdots,|\psi_{N}\rangle)^{\top}$ obey the Schr\"{o}dinger equation $i\hbar\frac{\partial}{\partial {t}}|\Psi\rangle =\tilde{H}_{\text{eff}}|\Psi\rangle $, with the effective Hamiltonian $\tilde{H}_{\text{eff}}=UHU^{\dagger}$ taking the following form:
\begin{equation}\label{H_eff}
\tilde{H}_{\text{eff}}=\frac{1}{2m}(-i\hbar \nabla -\mathbf{A})^{2}+\varepsilon I_N+\tilde{V}(\mathbf{r}).
\end{equation}
Here $\mathbf{A}=i\hbar U\nabla U^{\dagger }$, $\varepsilon=(\varepsilon_1,\varepsilon_2,\cdots,\varepsilon_N)^{\top}$, $\tilde{V}(\mathbf{r})=UV(\mathbf{r})U^{\dagger }$
and $I_N$ is the $N\times N$ unit matrix.  From Eq. (\ref{H_eff}), one can see that in the dressed basis the atoms can be considered as moving in an induced (artificial) vector potential $\mathbf{{A}}$ and a scalar potential $\tilde{V}(\mathbf{r})$, where the potential $\mathbf{A}$ is usually called the Mead-Berry vector potential. They come from the spatial dependence of the atomic dressed states with the elements $\mathbf{A}_{mn} = i\hbar \langle \chi_m(\mathbf{r})|\nabla\chi_n(\mathbf{r}) \rangle$, $\tilde{V}_{mn} = \langle\chi_m(\mathbf{r})|V(\mathbf{r})|\chi_n(\mathbf{r})\rangle$.

An artificial non-Abelian gauge field  can be induced in this way if there are degenerate (or nearly degenerate) dressed states. 
 Assume that the first $q$ atomic dressed states among the total $N$ states are degenerate, and these levels are well separated from the remaining $N-q$ states, we neglect the transitions from the first $q$ atomic dressed states to the remaining states. In this way, we can project the full Hamiltonian onto this subspace. Under this condition, the wave function in the subspace $\Psi_q=\left(\psi_1,\dots,\psi_q\right)^\top$ is again governed by the Schr\"odinger equation $i\hbar\frac{\partial}{\partial t}\Psi_q={H}_{\text{eff}}\Psi_q$, where the effective Hamiltonian reads
\begin{equation}\label{H_non}
{H}_{\text{eff}}=\frac{1}{2m}(-i\hbar\nabla -\mathbf{A})^2 +\varepsilon I_q + V_\text{eff}
\end{equation}
Here $ V_\text{eff}=\tilde{V}+ \tilde{V}^\prime$ and  the matrices $\mathbf{A}$, $\varepsilon I_q$, and $\tilde{V}$ are the truncated $q\times q$ matrices in Eq. (\ref{H_eff}). The projection of the term $\mathbf{A}^2$ in Eq. (\ref{H_eff}) to the $q$ dimensional subspace cannot entirely be expressed in terms of the truncated matrix $\mathbf{A}$. This gives rise to an additional scalar potential $\tilde{V}^\prime$ which is also a $q\times q$ matrix, $\tilde{V}^\prime _{n,j}=\frac{1}{2m}\sum_{l=q+1}^{N}\mathbf{A}_{n,l}\cdot\mathbf{A}_{l,j}$ with $n,j \in(1,\dots,q)$. Since the adiabatic states $|\chi_1\rangle \dots |\chi_q\rangle$ are degenerate, any basis generated by a local unitary transformation $U(\mathbf{r})$ within the subspace is equivalent. The corresponding local basis change as $\Psi \rightarrow U(\mathbf{r})\Psi,$ which leads to a transformation of the potentials according to
\begin{equation}
\mathbf{A} \rightarrow
U(\mathbf{r}) \mathbf{A}U^{\dag}(\mathbf{r})-i\hbar\left[\nabla U(\mathbf{r})\right] U^{\dag}(\mathbf{r}),
\end{equation}
and ${V}\rightarrow U(\mathbf{r}){V} U^{\dag}(\mathbf{r})$. These transformation rules show the gauge character of the potentials $\mathbf{A}$ and ${V}$. The vector potential $\mathbf{A}$ is related to a curvature (an effective ``magnetic'' field) $\mathbf{B}$ as:
\begin{equation}\label{eq:B}
B_i = \frac{1}{2}\epsilon_{ikl} F_{kl},~~ F_{kl}
=\partial_k A_l-\partial_l A_k-\frac{i}{\hbar}[A_k,A_l].
\end{equation}
Note that the term $\frac{1}{2}\varepsilon_{ikl}[A_k,A_l] =(\mathbf{A}\times \mathbf{A})_i$ does not vanish in general, since the components of $\mathbf{A}$ do not necessarily commute. This term reflects the non-Abelian character of the gauge potentials. The generalized ``magnetic" field transforms under local rotations of the degenerate dressed basis as $ \mathbf{B}\rightarrow U(\mathbf{r})\mathbf{B} U^{\dag}(\mathbf{r}).$ Thus, as expected, $\mathbf{B}$ is a true gauge field. The contents in this section can be found in Ref. \cite{DWZhang2018}, we repeat here just for convenience and completeness.

\subsection{Realizing synthetic topological vacua of Yang-Mills field}

\emph{Topological vacua in an artificial Yang-Mills gauge field}. An artificial $SU(2)$ Yang-Mills gauge field can be realized with  an atomic system which has two degenerate dressed states. We can well design the laser-atom interaction to lead the transform matrix $U$ in Eq.(\ref{Ua}) satisfying
\begin{equation}\label{U=Uw}
U(\mathbf{r})=U_n (\mathbf{r}).
\end{equation}
Furthermore, the scalar fields ${V}+{V}^\prime$ can be neglected in certain conditions or compensated with additional laser beams. Under these conditions, we can create a topological vacuum of the Yang-Mills field with the topological number $n$. Two completely degenerate dressed states appear in four-level tripod or cyclic configuration \cite{DWZhang2018}, and thus Yang-Mills gauge field can be induced in such systems. Notably, two nearly degenerate dressed states can be found in a large-detuning three-level $\Lambda$-type atomic system, and the non-Abelian gauge field can be  more easily realized in this three-level system, as we will show in the following.

\emph{Realization of topological vacua with a large-detuning three-level $\Lambda$-type atomic system}. We consider the induced gauge fields of an atomic gas with each atom having a $\Lambda $-type level configuration. As shown in Fig. 1 in the main text, the ground states $|1\rangle $ and $|2\rangle $ are coupled to an excited state $|e\rangle $ through spatially varying laser fields, with the corresponding Rabi frequencies $\Omega _{1}(\mathbf{r})$ and $\Omega _{2}(\mathbf{r})$, respectively. In addition, the coupling between states $|1\rangle$ and $|2\rangle$ is induced by a microwave and denoted as $\Omega_{MW}(\mathbf{r})$. We denote the single (two)-photon detuning as $\delta (\mathbf{r})$ ($\Delta (\mathbf{r})$). In this case
the atom-laser interaction Hamiltonian $H_{\text{AL}}$  in the basis $\left\{ |1\rangle ,|2\rangle ,|e\rangle \right\} $ is given by
\begin{equation}\label{H_int_3}
{H}_{\text{AL}}(\mathbf{r})=\frac{\hbar}{2}
\left(\begin{array}{cccc}
0& \Omega^*_{MW}(\mathbf{r}) & \Omega^*_1(\mathbf{r}) &\\
\Omega_{MW}(\mathbf{r}) & -\alpha(\mathbf{r})|\delta(\mathbf{r})| & \Omega^*_2(\mathbf{r}) &\\
\Omega_1(\mathbf{r}) & \Omega_2(\mathbf{r}) & \alpha(\mathbf{r})|\Delta(\mathbf{r})| &
\end{array}\right),
\end{equation}
where $\alpha(\mathbf{r})=sign(|\mathbf{r}|-1)$.
Just for easily obtaining a solution of Eq. (\ref{U=Uw}), we assume $\Omega_{MW}(\mathbf{r})=-i\Omega_1(\mathbf{r})\sqrt{|\delta(\mathbf{r})/\Delta(\mathbf{r})|}$. For simplifying the notations, we hide the notation $\mathbf{r}$ later on. We solve the Schrodinger equation $H_{\text{AL}}|\chi_n\rangle=\varepsilon_n|\chi_n\rangle$ under the large detuning conditions, i.e., $\Delta \gg |\delta|,\Omega$ with $\Omega=\sqrt{|\Omega_1|^2+|\Omega_2|^2}$ , and then obtain the eigenvalues $\varepsilon \approx 0,-\hbar\kappa^2/(4\Delta), \hbar[\Delta+\Omega^2/(4\Delta)]$ . The first two have large gap with the last one and can be considered as two nearly degenerate states, and thus they consist of a subspace with artificial Yang-Mills gauge field. After solving the related eigenvalues $\{\chi_1,\chi_2\}$ we can derive  the transform matrix in Eq.(5) in the main text.
Therefore, if we can find a solution with $U_{AL}=U_n$, then a topological vacuum with the topological  number $n$ is realizable.
We may parameterize the Rabi frequencies through $\Omega _{1} =-\Omega \sin \theta e^{i\varphi }$ and $ \Omega _{2} = \Omega \cos\theta$, with $\Omega =\sqrt{|\Omega _{1}|^{2}+|\Omega _{2}|^{2}}$.
\begin{equation}\label{U_para}
U_{\text{AL}}=\frac{2\alpha\sqrt{|\delta\Delta|}}{\kappa}I-\frac{i\Omega}{\kappa}
\left(\begin{array}{cc}
\cos\theta & \sin\theta e^{-i\varphi} \\
\sin\theta e^{i\varphi} & -\cos\theta
\end{array}\right).
\end{equation}
Comparing Eq.(\ref{H_original}) represented in the spherical coordinate system in the main text, if we choose
\begin{equation}
\Omega=2\sqrt{|\delta\Delta|}\tan[2\arctan(r/\eta)],
\end{equation}
then we obtain $U_{\text{AL}}=U_{1}$.

\begin{figure*}[htbp]\includegraphics[height=5cm]{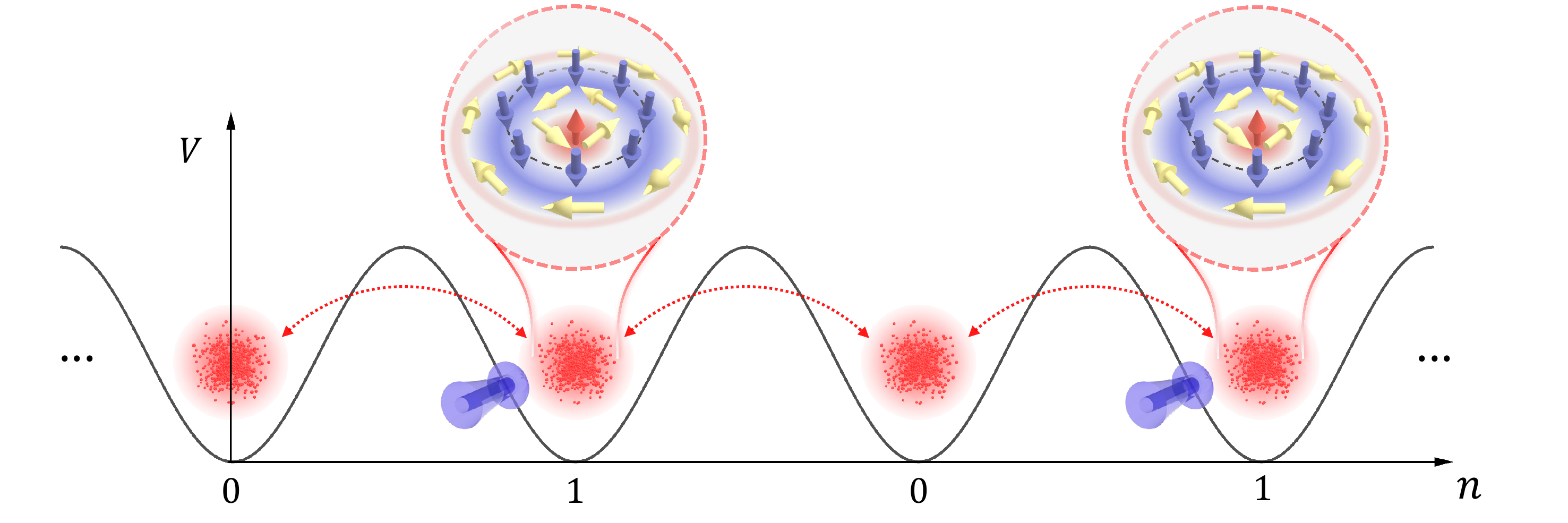}
\caption{Schematics of topological vacua array with period $n=1$ and $n=0$, which is the simplest configuration to have long-sought instantons. }
\label{fig:topvactrans}
\end{figure*}

We can further realize vacua with different TN $n$ in an array configuration as shown in Fig.1a in the main text, and then instantons can emerge in such an array. The simplest configuration for instantons  should be an array with alternating   $n=0$ and $n=1$ synthetic vacua, as shown in Fig.\ref{fig:topvactrans}. From spin textures plotted in Fig.2 in the main text, we find that the spin distribution around $r=4\eta$ for $n=1$ is almost the same with that at $r\rightarrow \infty$. Therefore, we can obtain a good topological vacuum array if the size of each cloud of atoms  in Fig.\ref{fig:topvactrans}  is larger than $4\eta$.

\subsection{Topological vacua and topological number}
If only gauge fields are present, the Yang-Mills Lagrangian is given by
\begin{equation}
\mathcal{L}_{YM}=-\frac{1}{2}tr\{F^{\mu\nu}F_{\mu\nu}\}=\frac{1}{2}(\mathbf{E}^2-\mathbf{B}^2),
\end{equation}
where $ {F}_{\mu\nu}$ is the field strength as defined in the main text.
The definition of electric and magnetic fields in terms of the field strength tensor is the same as that in electrodynamics, i.e., $E^{ia}=-F^{0ia}$, $B^{ia}=-\frac{1}{2}\epsilon^{ijk}F^{jka}$, here $\epsilon_{ijk}$ is the Levi-Civita symbol with $i,j,k \in \{x,y,z\}$ . As for $U_n=(U_1)^n$ defined in Eq. (1) in the main text, one can easily check that $\mathbf{E}=0$ and $\mathbf{B}=0$ since it is a pure gauge $\mathbf{A}=iU_n\nabla U_n^{-1}$. In $SU(2)$ Yang-Mills theory, any point $\mathbf{r}$  defines a $U(\mathbf{r}) \in SU(2)$. Each $U(\mathbf{r})$ defines a mapping from the base space $\mathcal{R}^3$ to $S^3$ since topologically $SU(2)\sim S^3$.  Furthermore, $U(\mathbf{r})$ defined in Eq. (1) in the main text approaches a unique value independent of the direction of $\mathbf{r}$: $U(\mathbf{r}) \rightarrow \text{const.}$ for $|\mathbf{r}|\rightarrow \infty $. So the configuration space is compact $\mathcal{R}^3 \rightarrow S^3$ and $U(\mathbf{r})$ define a map
\begin{equation}U(\mathbf{r}): \ \ S^3\rightarrow S^3.\end{equation}
Since the homotopy group $\pi_n(S^n)\sim \mathcal{Z}$, a winding number can be assigned for the mapping $U(\mathbf{r})$. This winding number counts how many times the three-sphere of gauge transformations $U(\mathbf{r})$ is covered if $\mathbf{r}$ covers once the three-sphere of the compactified configuration space.
The winding number $n$ can be expressed as
\begin{equation}\label{Windingnumber}
n=\frac{1}{24\pi^2}\int_{-\infty}^{+\infty} d\mathbf{r}\epsilon_{ijk}\text{Tr}( {U}^{-1}\partial_{i} {U} {U}^{-1}\partial_{j}  {U} {U}^{-1}\partial_{k} {U}).
\end{equation}
Here  the integration is done in the full space where the field lives.

Any $2\times 2$ matrix can be expanded in the basis $\sigma_0$ and  $\langle {\mathbf{\sigma}}\rangle=(\sigma_x,\sigma_y,\sigma_z)$ and then $U$ can be written as  $U=d_0\sigma^0+id_j\sigma^j$. The experimentally realized Hamiltonian ${H}=U\sigma_z U^{-}$ can be rewriten as $\mathbf{n}\cdot{\langle {\mathbf{\sigma}}\rangle}$, where $\mathbf{n}(\mathbf{r})=(d_3-id_0, d_1+id_2){\langle {\mathbf{\sigma}}\rangle}(d_3-id_0, d_1+id_2)^{\top}$ is a unit vector with $\mathbf{d}=(d_0,d_1,d_2,d_3)$ being a unit vector in 4-dimensional (4D) Euclidean space $\mathcal{R}^4$ and thus maps the coordinates of $\mathbf{d}$ on a 3D spherical surface $S^3$ to the coordinates $\mathbf{r}=(x,y,z)$ of $\mathbf{n}$ on a $S^2$, so we have an intrinsic Hopf map $h:S^3\rightarrow{S^2}$.  As a consequence, the underlying structure of the Hamiltonian ${H}$ represents a composite map $S^3\longrightarrow{u}S^3\longrightarrow{h}S^2$ from the 3D real space to the target pseudo-spin space. The topological invariant $\chi_{Hopf}$ of ${H}$ termed as Hopf index (also Hopf charge or Hopf invariant) is thus identical to the winding number\cite{Moore2008,XXYuan2017}:
\begin{equation}\label{wdef1}
	n=\chi_{Hopf}=-\frac{1}{4\pi^2}\int_{R^3} d^3x \mathbf{a}\cdot{\mathbf{f}},
\end{equation}
where $\mathbf{f}$ is the Berry curvature defined as $f_i=\epsilon_{ijk}\langle\partial_{j}\psi_-|{\partial_{k}\psi_-\rangle}$, $\rangle{\psi_-}$ denotes one of the eigenstates of the Hamiltonian $H$. $\mathbf{a}$ is the associated Berry connection with $\nabla\times\mathbf{a}=\mathbf{f}$. This result can be explained by decomposing the composition map $\chi_{Hopf}(\mathbf{n})=\chi_{Hopf}(h)n(u)$, where the intrinsic Hopf map from $S^3$ to $S^2$ has a famous Hopf index $\chi_{Hopf}(h)=1$. The map $u:S^3\rightarrow S^3$ are classified by the winding number expressed in Eq. (\ref{Windingnumber}).

The geometry images of these topological vacua with Hopf index $\chi_{Hopf}=n$ are related to the Hopf links. Take the case $n=1$ for an example, a point such as the north pole in $S^2$ should be mapped to a closed circle in $\mathcal{R}^3$ and the interwinding times, termed as linking number, of two such circles mapped from different points of $S^2$ is just the Hopf index. For the vacua with higher topological numbers, a point in $S^2$ is mapped to $n$ concentric circles in $\mathcal{R}^3$. Two different points in $S^2$ have the linking numbers $n$ where each two circles belong to different preimages links once.

\subsection{Detection scheme of Hopf index}

With definition of the Hopf index in Eq.(\ref{wdef1}), it is easy to find that $\mathbf{f}$ is gauge invariant and thus experimentally measurable. However, $\mathbf{a}$ is gauge dependent and thus need to be derived from the experimental data with a proper gauge. This gauge dependence is actually from the undetectable global phase of the local wavefunction of atoms, e.g. $|\psi_\pm(\mathbf{r})\rangle$. For an atom, the most ordinary quantity we may experimentally obtain is the density matrix, which is gauge invariant and can be measured by atom population. To measure the winding number $n$ or Hopf index $\chi_{Hopf}$, we here use the  method proposed in Ref.\cite{XXYuan2017}. First, we measure the discretized Berry curvature $\mathbf{f}_{i}(\mathbf{r_J})$ and then calculate the Berry connection $\mathbf{a}_{i}(\mathbf{r_J})$ by solving a discrete version of the equation $\nabla\times\mathbf{a} =\mathbf{f}$ with certain gauge, e.g. the Coulomb gauge $\nabla\cdot\mathbf{a}=0$. Finally, we calculate the topological number $n$ ( the Hopf invariant $\chi$) by a discrete sum over all the points in a finite measurement space.

Here, we need a quantity that the so-called U(1)-link ${U}_i(\mathbf{r_J})=\langle\psi_-(\mathbf{r_J})|\psi_-(\mathbf{r}_{\mathbf{J}+\delta{i}})\rangle$ to show the details of measuring the discretized Berry curvature ${f}_{i}(\mathbf{r_J})$,
where $\delta i \in \{\delta x,\delta y,\delta z\}$ are infinitesimal vectors in the corresponding direction.
It is easy to see that ${U}_i(\mathbf{r_J})=1+{a}_{i}(\mathbf{r_J})\delta i \approx e^{{a}_{i}(\mathbf{r_J})\delta i}$,
where ${a}_{i}(\mathbf{r_J})=\langle\psi_-(\mathbf{r_J})|\partial_{i}|\psi_-(\mathbf{r_J})\rangle$.
The Berry curature is then
\begin{align}\label{Berrycurvature}
{f}_{i}(\mathbf{r_J})= & \left[ \partial_{j} {a}_{k}(\mathbf{r_J})-\partial_{k} {a}_{j}(\mathbf{r_J})\right]\delta j\delta k/\delta j\delta k\notag\\
     = & \frac{ \left[ {a}_{k}(\mathbf{r}_{\mathbf{J}+\delta {j}})\delta k- {a}_{k}(\mathbf{r_J})\delta k- {a}_{j}(\mathbf{r}_{\mathbf{J}+\delta {k}})\delta j+ {a}_{j}(\mathbf{r_J})\delta j\right]}{\delta j\delta k}\notag\\
    = & \left[\text{ln}{U}_{j}(\mathbf{r_J}){U}_{k}(\mathbf{r}_{\mathbf{J}+\delta {j}}){U}^{-1}_{j}(\mathbf{r}_{\mathbf{J}+\delta {k}}){U}^{-1}_{k}(\mathbf{r_J})\right]/\delta j\delta k.
\end{align}
Notice that $\langle\psi_1|\psi_2\rangle\langle\psi_2|\psi_3\rangle\langle\psi_3|\psi_4\rangle\langle\psi_4|\psi_1\rangle=\text{tr} (\rho_1\rho_2\rho_3\rho_4)$ with $\rho_i=|\psi_i\rangle\langle\psi_i|$, the discrete Berry curvature in Eq. (\ref{Berrycurvature}) can thus be expressed as
\begin{equation}\label{berrydensitymat}
	{f}_i(\mathbf{r_J})=\text{ln}~ \text{tr}[\rho(\mathbf{r_J})\rho(\mathbf{r}_{\mathbf{J}+\delta {j}})\rho(\mathbf{r}_{\mathbf{J}+\delta {j}+\delta {k}})\rho(\mathbf{r}_{\mathbf{J}+\delta {k}})]/\delta j\delta k.
\end{equation}
Therefore, by choosing Coulomb gauge, we can derive Berry connection from Berry curature in the following way:
\begin{equation}
	\mathbf{a}(\mathbf{r})=\int_{V'}\frac{\nabla\times{\mathbf{f}(\mathbf{r}')}dV'}{4\pi\left|\mathbf{r}-\mathbf{r}'\right| }.
\end{equation}
Since $\mathbf{f}(\mathbf{r}')=\int_{V}\mathbf{f}(\mathbf{k})e^{-i\mathbf{k}\cdot\mathbf{r}'}d^3k$,
\begin{equation}
    \int_{V'}\frac{e^{i\mathbf{k}\cdot(\mathbf{r}-\mathbf{r}')}dV'}{4\pi\left|\mathbf{r}-\mathbf{r}'\right| }=\frac{1}{k^2},
\end{equation}
so that
\begin{equation}
	\mathbf{a}(\mathbf{r})=\int_{V}\frac{-i\mathbf{k}\times{\mathbf{f}(\mathbf{k})}}{k^2}e^{-i\mathbf{k}\cdot\mathbf{r}}d^3k=\int_{V}\mathbf{a}(\mathbf{k})e^{-i\mathbf{k}\cdot\mathbf{r}}d^3k,
\end{equation}
with $\mathbf{a}(\mathbf{k})=\frac{-i\mathbf{k}\times{\mathbf{f}(\mathbf{k})}}{k^2}$.
For discrete case, we calculate ${\mathbf{f}}(\mathbf{k_J})$, ${\mathbf{a}}(\mathbf{k_J})$ and ${\mathbf{a}}(\mathbf{r_J})$ in turn by discrete Fourier transform.
Finally we get
\begin{equation}\label{wnumsum}
	\chi_{Hopf}=-\frac{1}{4\pi^2}\sum_{\mathbf{r_J}}{\mathbf{f}}(\mathbf{r_J})\cdot{\mathbf{a}}(\mathbf{r_J}){\Delta\mathbf{r}}^3
\end{equation}
Experimentally, density matrix $\rho_{\mathbf{r}}$ at point $\mathbf{r}$ are measured by quantum state tomography, then the topological invariants can be probed with the above method.

\section{Experimental scheme}
Multilevel atoms with three selected internal energy levels denoted as $\{|1\rangle,|2\rangle,|e\rangle \}$  are used to realize the Hamiltoninan described in Eq. (4) in the main text and Eq.(\ref{H_int_3}), where the three energy levels are cyclically coupled by a pair of Raman lasers and a microwave, as shown in Fig.1 in the main text.
\subsection{Effective Hamiltonian of lasers and atoms}
In the semiclassical region that laser fields are intense and  thus can be treated as classical field, the Hamiltonian of the atomic system is given as
\begin{equation}
{H}_{\text{AL0}}=H_0+H_{\text{int}},
\end{equation}
where $H_0$ denotes the Hamiltonian of electron and nuclear inside the atom that is a diagonal $N\times N$ matrix in its own eigenstate basis of the $N$ internal energy levels, such as $\{|1\rangle,|2\rangle,|e\rangle \}$ in our work. The laser-atom interaction $H_{\text{int}}$ depends on the position of the atoms and is typically a non-diagonal $N\times N$ matrix in the above basis. In our work, the three basis states are coupled by three electromagnetic (EM) waves $\mathbf{E}_l=\frac{1}{2}E_l e^{(i \mathbf{k}_l \mathbf{r} -i\omega_l t)}+c.c.$ with $E_l$ being the EM wave amplitude, $c.c.$ meaning complex conjugate and index $l=\{1,2,MW\}$.

Considering that the size of an atom is on the order of $0.1~nm$ while the EM waves in our work have wavelength much larger (on order of $\mu m$ and  $cm$), it is safe to take the commonly used dipole approximation that an atom would feel spatially homogeneous field. The fast oscillating EM waves would change the electron wave function inside the atom and induce electric or magnetic dipole moment with its matrix element $\mu_{mn}$ in the above eigenstate basis. The induced dipoles further interact with these external EM fields and we usually use the Rabi frequency to  describe their interaction strength. In our work, Rabi frequency corresponding to two laser fields are defined as $\Omega_{1}=\mu_{1e}E_1/\hbar=-eE_1\langle 1|\mathbf{r}|e\rangle/\hbar$ and $\Omega_{2}=\mu_{2e}E_2/\hbar=-eE_2\langle 2|\mathbf{r}|e\rangle/\hbar$. The microwave couples $\{|1\rangle,|2\rangle \}$ via magnetic dipole interaction and hence the Rabi frequency is $\Omega_{MW}=\mu_{12}B_{MW}/\hbar=B_{MW}g\langle 2|\mathbf{F}|1\rangle/\hbar$ with $g$ being the gyromagnetic ratio of the transition and $\mathbf{F}$ being the total angular momentum operator of the atom. Proper polarization of EM waves are chosen in experiment to ensure that the coupling is allowed by the selection rules. As the interaction of an atom with an EM wave shows a significant resonant line-shape, we may consider only the most resonant coupling in the following derivation. Therefore, the Hamiltonian ${H}_{\text{AL0}}$ reads
\begin{widetext}
\begin{equation}\begin{array}{ll}
{H}_{0}=\hbar\omega^{atom}_1|1\rangle\langle1|+\hbar\omega^{atom}_2|2\rangle\langle2|+\hbar\omega^{atom}_e|e\rangle\langle e|,\\
H_{\text{int}}=\frac{1}{2}(\hbar\Omega^*_{MW}e^{-i\omega_{MW}t}+c.c)|1\rangle\langle2|+\frac{1}{2}(\hbar\Omega_{1}e^{-i\omega_l t}+c.c)|1\rangle\langle e|+\frac{1}{2}(\hbar\Omega_{2}e^{-i\omega_2 t}+c.c)|2\rangle\langle e|+h.c.,
\end{array}\end{equation}
\end{widetext}
where $h.c.$ means the Hermitian conjugate. ${H}_{AL0}$ can be rewriten to a $3\times 3$ matrix:
\begin{widetext}
\begin{equation}
{H}_{\text{AL0}}=\hbar
\left(\begin{array}{cccc}
\omega^{atom}_1& \frac{1}{2}\Omega^*_{MW}e^{-i\omega_{MW}t}+c.c. & \frac{1}{2}\Omega_{1}e^{-i\omega_1 t}+c.c.&\\
\frac{1}{2}\Omega_{MW}e^{i\omega_{MW} t}+c.c. & \omega^{atom}_2 & \frac{1}{2}\Omega_{2}e^{-i\omega_2 t}+c.c.&\\
\frac{1}{2}\Omega^*_{1}e^{i\omega_1 t}+c.c. & \frac{1}{2}\Omega^*_{2}e^{i\omega_2 t}+c.c. & \omega^{atom}_e &
\end{array}\right),
\end{equation}
\end{widetext}
where $\hbar\omega^{atom}_m$ denote the energy of eigenstate $\{|1\rangle,|2\rangle,|e\rangle \}$ with $m=\{1,2,e\}$.

The above Hamiltonian is obviously time-dependent and thus we may try to implement a time dependent transformation and investigate the problem in a rotating frame. The transformation matrix is assumed to take the form $\hat{U}=e^{\frac{i}{\hbar}\Lambda t}$, where  $\Lambda$ takes the following form
\begin{equation}
\Lambda=\hbar
\left(\begin{array}{cccc}
\lambda_1 & 0 & 0 \\
0 & \lambda_2 & 0 \\
0 & 0 & \lambda_3
\end{array}\right).
\end{equation}
Then we may have the effective Hamiltonian $H_{\text{AL}}=\hat{U}H_{\text{AL0}}\hat{U}^\dagger-\Lambda$ which  satisfies the Schr\"odinger equation $H_{\text{AL}}|\chi_n\rangle=\varepsilon_n|\chi_n\rangle$ in the rotating frame:
\begin{widetext}
\begin{equation}
{H}_{\text{AL}}=\hbar
\left(\begin{array}{cccc}
\omega^{atom}_1-\lambda_1
&\frac{1}{2}(\Omega^*_{MW}e^{-i\omega_{MW}t}+c.c.)e^{-i(\lambda_2-\lambda_1)t}
& \frac{1}{2}(\Omega_{1}e^{-i\omega_1 t}+c.c.)e^{-i(\lambda_3-\lambda_1)t}&\\

\frac{1}{2}(\Omega_{MW}e^{i\omega_{MW} t}+c.c.)e^{i(\lambda_2-\lambda_1)t}
& \omega^{atom}_2-\lambda_2
& \frac{1}{2}(\Omega_{2}e^{-i\omega_2 t}+c.c.)e^{-i(\lambda_3-\lambda_2)t}&\\

\frac{1}{2}(\Omega^*_{1}e^{i\omega_1 t}+c.c.)e^{i(\lambda_3-\lambda_1)t}
& \frac{1}{2}(\Omega^*_{2}e^{i\omega_2 t}+c.c.)e^{i(\lambda_3-\lambda_2)t}
& \omega^{atom}_e-\lambda_3 &
\end{array}\right).
\end{equation}
\end{widetext}

By taking $\lambda_1=\omega^{atom}_1$, $\lambda_3-\lambda_1=\omega_1$, $\lambda_3-\lambda_2=\omega_2$, and $\omega_{MW}=\omega_2-\omega_1=-(\lambda_2-\lambda_1)$, we have
\begin{widetext}
\begin{equation}
{H}_{\text{AL}}=\hbar
\left(\begin{array}{cccc}
0
&\frac{1}{2}(\Omega^*_{MW}+\Omega_{MW}e^{i2\omega_{MW}t})
& \frac{1}{2}(\Omega_{1}e^{-i2\omega_1 t}+\Omega^*_{1})&\\

\frac{1}{2}(\Omega^*_{MW}e^{-i2\omega_{MW}t}+\Omega_{MW})
& \omega^{atom}_2-(\omega^{atom}_1+\omega_1-\omega_2)
& \frac{1}{2}(\Omega_{2}e^{-i2\omega_2 t}+\Omega^*_{2})&\\

\frac{1}{2}(\Omega^*_{1}e^{i2\omega_1 t}+\Omega_{1})
& \frac{1}{2}(\Omega^*_{2}e^{i2\omega_2 t}+\Omega_{2})
& \omega^{atom}_e-(\omega^{atom}_1+\omega_1) &
\end{array}\right).
\end{equation}
\end{widetext}
As the frequencies of EM waves $\omega_{1,2,MW}$ are on the order of $10^{14} Hz$ and $10^{9} Hz$ and they are much faster than the typical physical procedure in our work, it is safe to further take the rotating wave approximation that neglects those terms oscillating in frequency of $2\omega_{1,2,MW}$. We further define the single photon detuning $\Delta=(\omega^{atom}_e-\omega^{atom}_1)-\omega_1$ and two-photon detuning $\delta=(\omega^{atom}_1-\omega^{atom}_2)-(\omega_2-\omega_1)$, we eventually have
\begin{equation}\label{H_int_3}
{H}_{\text{AL}}= \frac{\hbar}{2}
\left(\begin{array}{cccc}
0 &\Omega^*_{MW} &\Omega^*_{1}&\\
\Omega_{MW} & -2\delta & \Omega^*_{2}&\\
\Omega_{1} &  \Omega_{2} & 2\Delta &
\end{array}\right).
\end{equation}

By defining $\varepsilon=\frac{\hbar}{2}\epsilon$, the eigenvalues of Hamiltonian Eq.(\ref{H_int_3}) can be obtained from the below equation:
\begin{widetext}
\begin{equation}
-\epsilon^3+2(\Delta-\delta)\epsilon^2+ \epsilon(|\Omega_{1}|^2+|\Omega_{2}|^2-|\Omega_{MW}|^2+4\delta\Delta)+2 Re[\Omega_{MW}\Omega_{2}\Omega^*_{1}]+2\delta|\Omega_{1}|^2-2\Delta|\Omega_{MW}|^2=0.
\end{equation}
\end{widetext}
We here consider the large detuning condition $|\Delta|\gg|\delta|,\Omega$ with $\Omega=\sqrt{|\Omega_1|^2+|\Omega_2|^2}$. $\Omega_{2}$ could be set as a real number in experiment by choosing the initial phase of $\mathbf{E_2}$ as zero. Then by taking $\Omega_{MW}=-i\Omega_1\sqrt{|\delta/\Delta|}$, we have:
\begin{equation}
-\epsilon^3+2(\Delta-\delta)\epsilon^2+ \epsilon(|\Omega|^2-|\Omega_{MW}|^2+4\delta\Delta)=0.
\end{equation}
The eigenvalues are obviously $\epsilon = 0,(\Delta-\delta)\pm\sqrt{(\Delta+\delta)^2+\Omega^2-|\Omega_1|^2|\delta/\Delta|}$. Using the large detuning condition and the approximation $\sqrt{1+x}\approx1+x/2$ when $|x|\ll1$, we may further have the approximated egienvalues $\epsilon  \approx 0,-\kappa^2/(2\Delta), 2\Delta+\Omega^2/(2\Delta)$ with $\kappa=\sqrt{\Omega^2+4|\delta\Delta|},$ as shown in the main text.
 The state with $\epsilon=2\Delta+\Omega^2/(2\Delta)$ has an energy much far away from the remained two states and thus is effectively isolated,  while the left two eigenstates $\{|\chi_1\rangle,|\chi_2\rangle\}$ form a pseudospin subspace, and we denote them as $\{|\chi_{-}\rangle,|\chi_{+}\rangle\}$. The eigenvectors can be solved from the following  equations
\begin{equation}
\left(\begin{array}{cccc}
0 &i\Omega^*_1\sqrt{|\delta/\Delta|} &\Omega^*_{1}&\\
-i\Omega_1\sqrt{|\delta/\Delta|} & -2\delta & \Omega_{2}&\\
\Omega_{1} &  \Omega_{2} & 2\Delta &
\end{array}\right)
\left(\begin{array}{c}a\\b\\c\end{array}\right)
=\epsilon
\left(\begin{array}{c}a\\b\\c\end{array}\right).
\end{equation}
It would be straightforward to obtain the eigenvector for $\epsilon_-=\epsilon_1=0$, i.e.,
\begin{widetext}
\begin{equation}
|\chi_1\rangle=|\chi_{-}\rangle=(-i\Omega_2-2\Delta\sqrt{|\delta/\Delta|}, i\Omega_1, \Omega_1\sqrt{|\delta/\Delta|})^T/\sqrt{4|\delta\Delta|+\Omega^2+|\Omega_1|^2|\delta/\Delta|}.
\end{equation}
\end{widetext}
Under the large detuning approximation, we may safely ignore the component $\Omega_1\sqrt{|\delta/\Delta|}$ and thus  $|\chi_{-}\rangle$ is given as
\begin{equation}
|\chi_-\rangle=(-i\Omega_2-2\Delta\sqrt{|\delta/\Delta|}, i\Omega_1)^T/\kappa.
\end{equation}
Considering the orthognality between $|\chi_{-}\rangle$ and $|\chi_{+}\rangle$ in the pseudospin subspace, we can obtain the eigenvector $|\chi_{+}\rangle$ for $\epsilon_+\approx-2\delta-\Omega^2/(2\Delta)$,
\begin{equation}
|\chi_{+}\rangle=(i\Omega^*_1, i\Omega_2-2\Delta\sqrt{|\delta/\Delta|})^T/\kappa.
\end{equation}
Eventually $U_{\text{AL}}=(|\chi_{-}\rangle,|\chi_{+}\rangle)$ can be derived as
\begin{equation}\label{U_para}
U_{\text{AL}}=\frac{1}{\kappa}
\left(\begin{array}{cc}
-i\Omega_2+2\alpha\sqrt{|\delta\Delta|} & i\Omega_1^* \\
i\Omega_1 & i\Omega_2+2\alpha\sqrt{|\delta\Delta|}
\end{array}\right).
\end{equation}
where the sign function $\alpha$ is defined in the main text and the possible sign of $\Delta$ can be absorbed into $\alpha$.

In order to implement this Hamiltonian in real space, both frequency and power of lasers and microwave need to be position dependent and can be engineered as shown below:
\begin{equation}
{H}_{\text{AL}}(\mathbf{r})= \frac{\hbar}{2}
\left(\begin{array}{cccc}
0 &\Omega_{MW}(\mathbf{r}) &\Omega^*_{1}(\mathbf{r})&\\
\Omega^*_{MW}(\mathbf{r}) & -2\alpha(\mathbf{r})|\delta(\mathbf{r})| & \Omega^*_{2}(\mathbf{r})&\\
\Omega_{1}(\mathbf{r}) &  \Omega_{2} & 2\alpha(\mathbf{r})|\Delta(\mathbf{r})| &
\end{array}\right),
\end{equation}
which is experimentally challenging but theoretically achievable by spatially varying magnetic fields via Zeeman effects or spatially varying detuned laser field via ac-Stark shift. The Rabi frequency $\Omega_{1/2/MW}$ can be engineered by tuning the laser power and microwave power while the spatial dependence can be possibly realized by designing the laser and microwave intensity pattern with spatial light modulators or optical beam scanning technique.

For realization of topological vacuum with winding numebr $n=1$, the exact Hamiltonian Eq.(\ref{H_int_3}) can be determined by substituting Eq.(1) of main text into Eq.(\ref{U_para}) and then obtaint the specific spatial dependence of $\Omega_{1/2/MW}(\mathbf{r})$, $\Delta(\mathbf{r})$ and $\delta(\mathbf{r})$ accordingly. Here, it is clear that we actually have enough freedom to choose proper strength of Rabi frequency $\Omega$ as only the ratio of $\Omega_{1/2/MW}(\mathbf{r})$, $\Delta(\mathbf{r})$ and $\delta(\mathbf{r})$ to $\Omega$ determines the structure of simulated gauge field.  On the other hand, the absolute beam size of laser/microwave field can also be freely chosen as the part matters is the ratio of $\mathbf{r}/\eta$.
By further checking, as the topology of simulated gauge field vacuum depending only on the structure of two lower eigenvectors of Hamiltonian Eq.(\ref{H_int_3}), we may freely tune their eigen energy dependence on position $\mathbf{r}$ of two lower states. Here, in possible experiment, in order to maitain the degeneracy condition, the absolute energy level difference of two lower eigenstates is usually required to be small enough and the remained third energy level is much higher. However, even with such many experimental knots, there is still no feasible way for implementing the exact Hamiltonian Eq.(\ref{H_int_3}) in real space.
Therefore, as an alternative way, we here realize the Hamiltonian Eq.(\ref{H_int_3}) in a parameter space while the position $\mathbf{r}$ is a  parameter and can be tuned by all the controllable experiment quantities.

Although our experiment is just implemented in a parameter space, it still makes valuable contributions:   the spatial structure and the Hopf links of the topological vacuum are visualized in our experiment, which have not been explored in previous literature.

Aiming to realizing an SU(2) non-abelian vacuum described by $U_n$, the corresponding 3-level interaction Hamiltonian of atoms is desired to have the two degenerate eigenstates consisting $U_{\text{AL}}=U_n$, where the rest higher energy eigenstate is ignorable with a large single photon detuning $\Omega/\Delta \approx 10^{-7}$. Meanwhile, this large $\Delta$ also maintains a long life time of ultracold atoms in real experiment. Therefore, we effectively manipulate a nearly degenerate two-level system in the following way:
\begin{equation}\label{Heff_int2_flatband}
{H}_{TV}\propto U_{\text{AL}}\sigma_z U^{-1}_{\text{AL}} \propto U_{n}\sigma_z U^{-1}_{n}.
\end{equation}
In a parameter space, we may directly realize this effective two level Hamiltonian by tuning the cyclic coupling strength and Raman two-photon detuning for each specified position $\mathbf{r}$ and then measuring the topological invariant like winding number and also demonstrate the respective topology. Considering that $\delta/\Delta \approx 10^{-7}$ and $\Omega_{1,2}$ are on the order of kHz, the strength of microwave $\Omega_{MW}$ in Eq.(\ref{H_int_3}) is expected to be around Hz and thus would be safely approximate to be zero in experiment. Eventually, the three level ultracold atoms are effectively only coupled by a pair of Raman lasers with large enough single photon detuning $\Delta$ and finite tunable two-photon detuning $\delta$. The realized Hamiltonian is usually given as the following form:
\begin{equation}\label{Hexp}
{H}_\text{eff}=\dfrac{\hbar}{2}
	\begin{pmatrix}
		\delta(\mathbf{r}) & \Omega_\text{eff}(\mathbf{r}) e^{-i\varphi_0(\mathbf{r})} \\
		\Omega_\text{eff}(\mathbf{r}) e^{i\varphi_0(\mathbf{r})} & -\delta(\mathbf{r})
	\end{pmatrix},
\end{equation}
where $\Omega_\text{eff} =\Omega_1 \Omega_2 / \Delta$ is the effective two-level Rabi frequency and $\delta$ is the Raman two-photon detuning. The topological vacuum Hamiltonian is realized by required that  $H_\text{TV}=H_\text{eff}$.  By adiabatically tuning these experiment parameters, we may produce the desired laser-atom interaction described by above Hamiltonian and load atoms into the spin states consisting of $U_n$. Eventually, the topological winding number $n$ and spatial spin texture in parameter space are revealed by measuring the realized state of atom with quantum state tomography, as detailed in the following sections.

\section{Quantum state preparation, control and measurement}
\begin{figure}[htbp]
{\includegraphics[height=5cm]{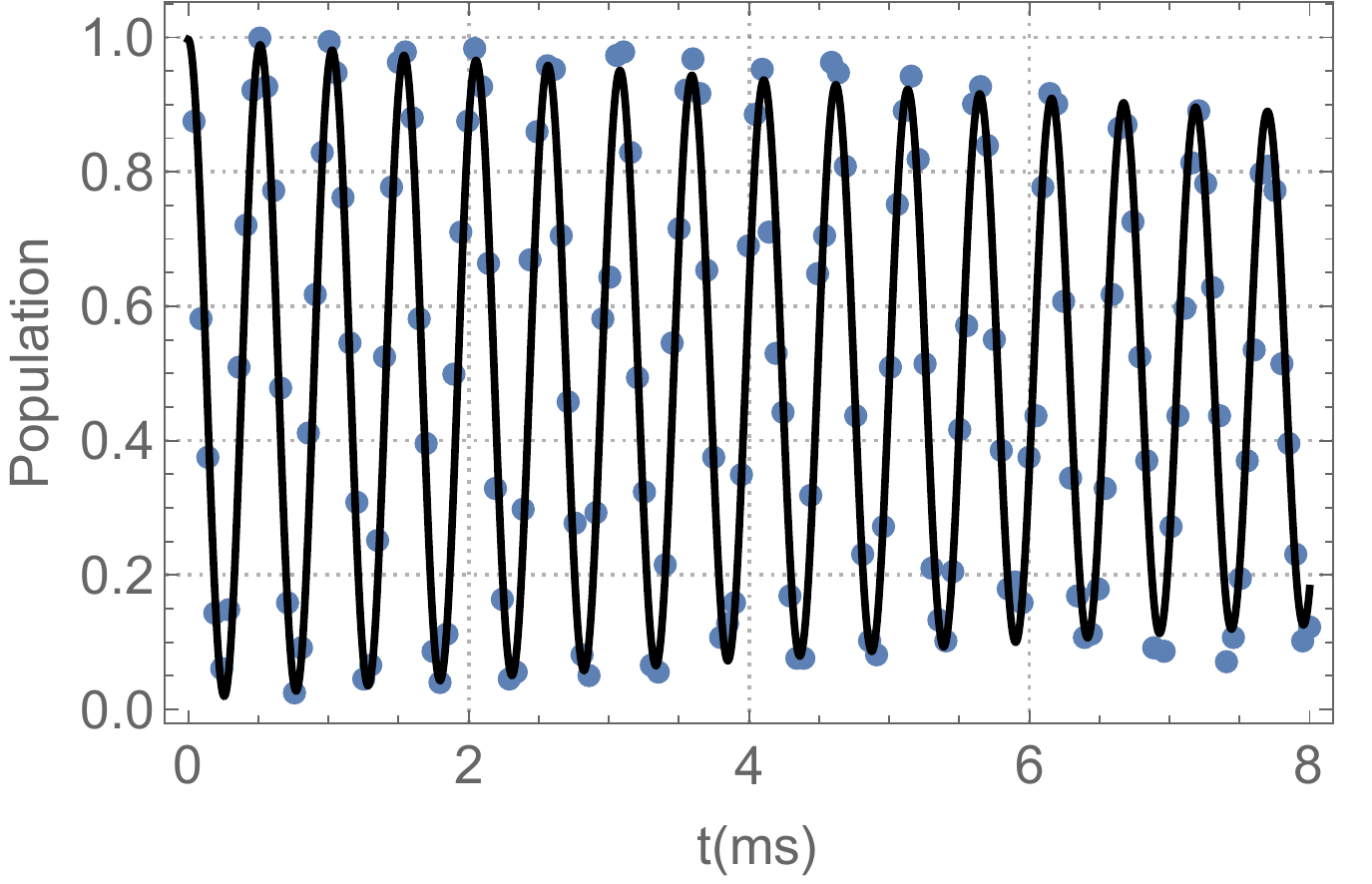}}
{\includegraphics[height=5cm]{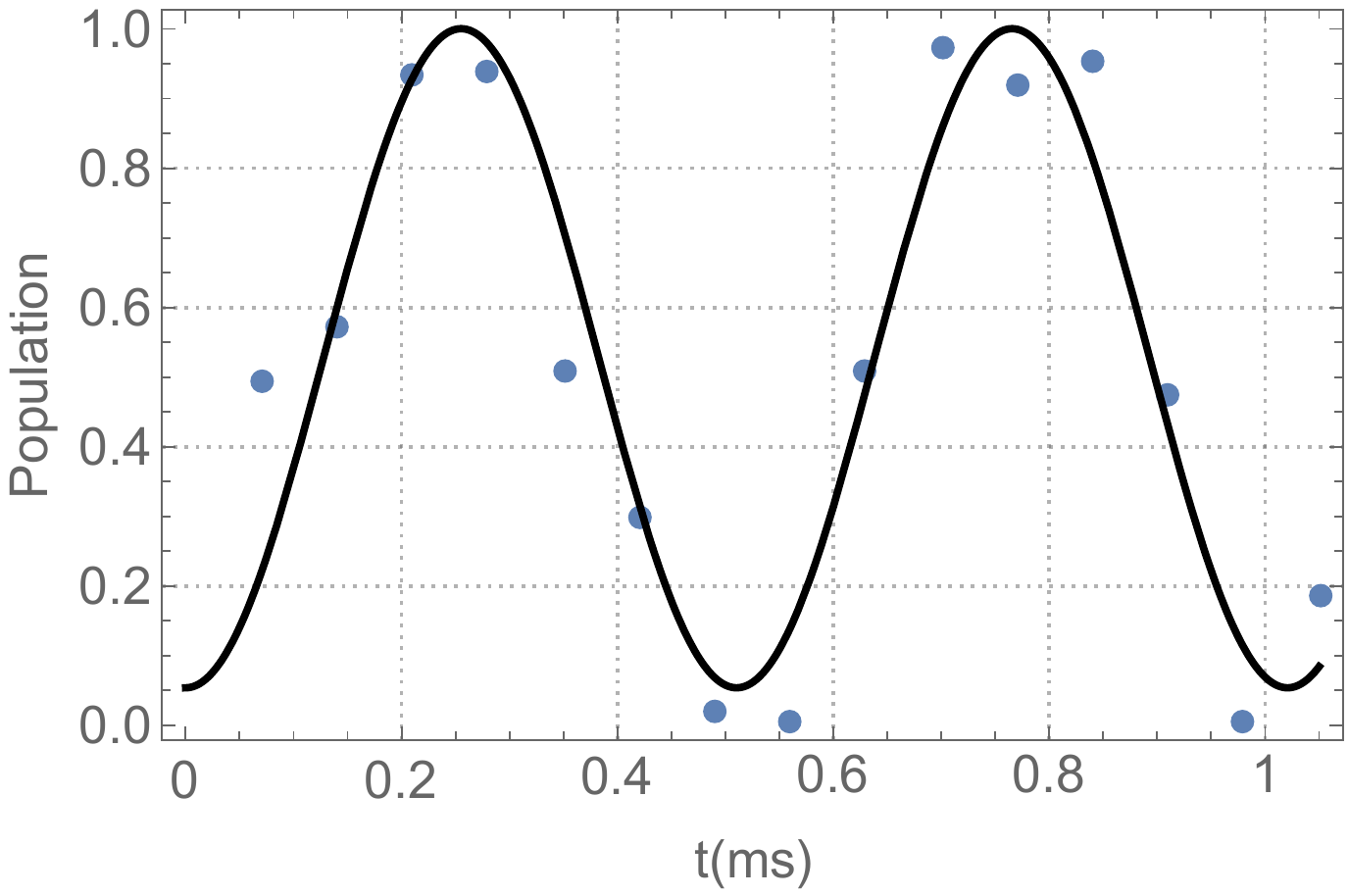}}
\caption{Coherence time characterization of transition between $\rangle{1}\leftrightarrow\rangle{2}$. (a) Ramsey fringe driven by resonant  microwave pulse shows coherence time $t_2$ longer than 8ms. (b) Ramsey fringe driven by two-photon resonant Raman laser pulse shows coherence time $t_2$ longer than 1ms.}
\label{fig:Ramsey}
\end{figure}
\subsection{Quantum state preparation and  measurement}
The BEC cloud contains about $ 2\times10^5$ atoms with temperature around $60\ nK$. Initially, all atoms are populated in the hyperfine Zeeman state of $|{1}\rangle=|{F=2,m_F=0}\rangle$ of the ground fine energy level $5S_{\frac{1}{2}}$ by using a microwave pulse resonant to the transition between ground hyperfine states $|{2}\rangle=|{F=1,m_F=0}\rangle\leftrightarrow|{1}\rangle$. A homogeneous quantized magnetic field around 3 Gs is generated by a pair of Helmholtz coils that set the quantum axis. Under this condition, the atoms show a coherent time of $T_2>8ms$ and $T_2>1ms$ for microwave and Raman coupling, as shown by the microwave driven Rabi oscillation and Ramsey fringe in Fig.\ref{fig:Ramsey}.

\begin{figure}[htbp]
{\includegraphics[height=4.5cm]{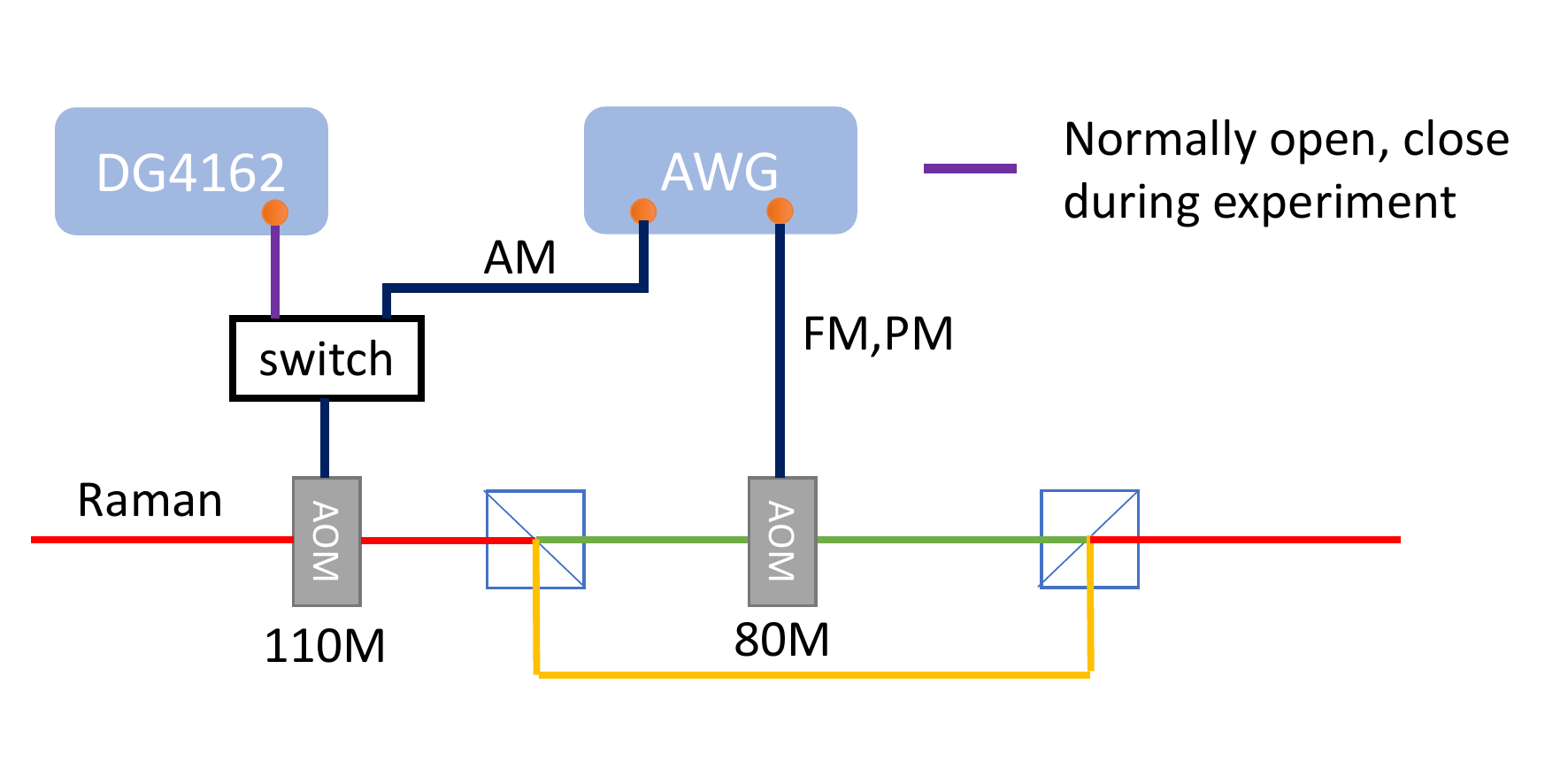}}
{\includegraphics[height=4.5cm]{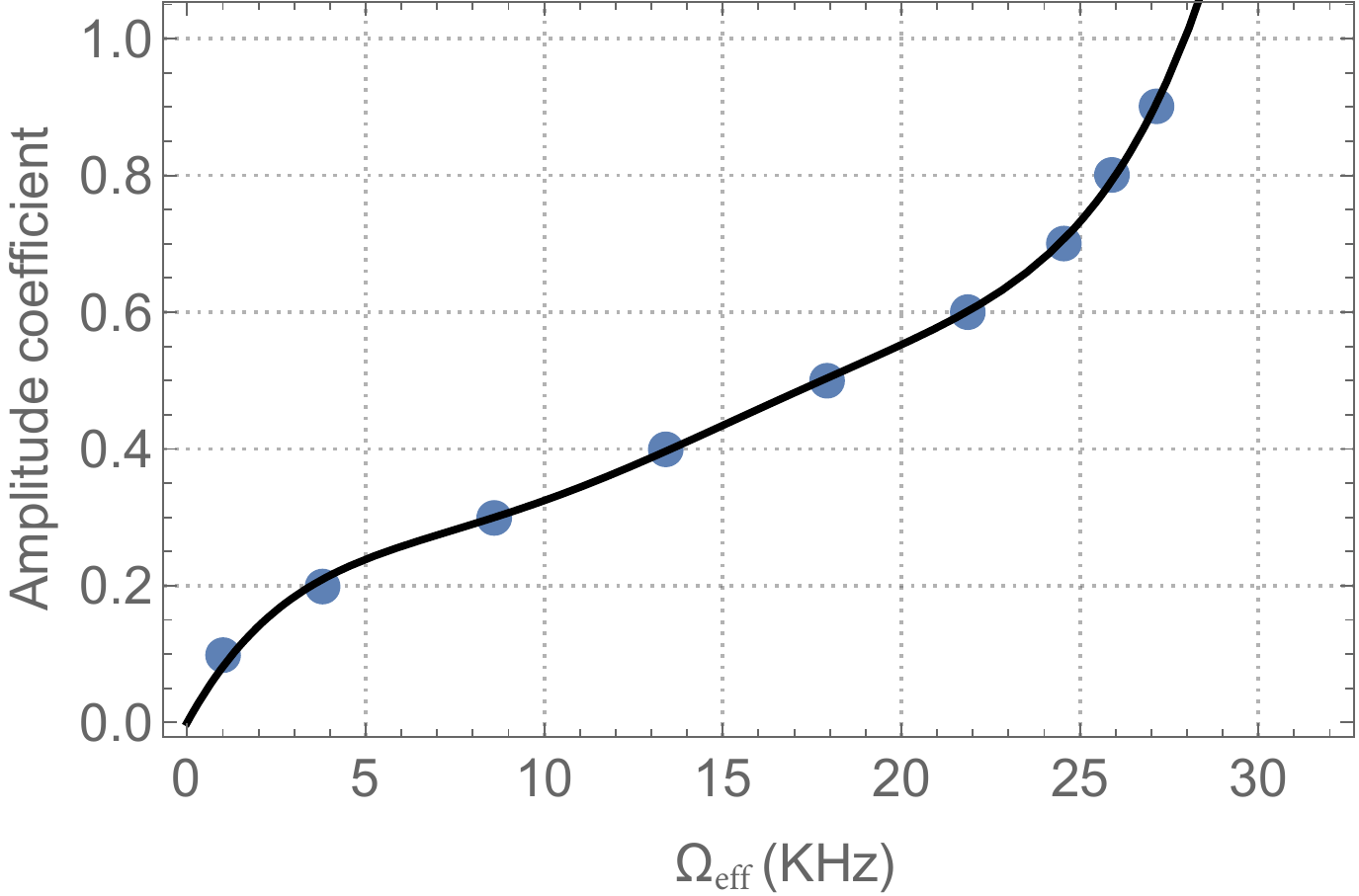}}
\caption{Experimental scheme for amplitude control and frequancy control of Ramam laser.(a) Schematics of the optical setup. The AOM working at 110MHz, which is shared by two Raman laser beams, is used to modulate the power of Raman laser and switch on or off the Raman lasers. While the AOM working at 80MHz is used to control the relative frequency difference of Raman lasers. (b) Experimentally obtained relation between the amplitued control voltage and the effective Raman Rabi frequency $\Omega_{eff}$. The solid line is a a 5th-order polynomial fit curve.}
\label{fig:OmegaVsAWGVpp}
\end{figure}

A pair of copropagating Raman laser beams are shined on the atoms in order to coherently couple states $|{1}\rangle$ and $|{2}\rangle$ via a two-photon Raman process. The frequency difference of Raman lasers is phase-locked to 6754.688082 MHz, which is set 80 MHz red detuned from the transition $|{1}\rangle\leftrightarrow|{2}\rangle$. The frequency of one laser beam is shifted through an acousto-optical modulator (AOM-80MHz) to meet experimentally required dynamical control of two-photon detuning $\delta$, as shown in Fig.\ref{fig:OmegaVsAWGVpp}. For the sake of lifetime of ultracold atoms in ODT, both Raman lasers have wavelength around 788nm and thus the single photon detuning of Raman lasers respect to the D2 transition of $^{87}Rb$ is esitmated to be $\Delta/(2\pi)=c/780 nm-c/788 nm= 3.9$ THz, which shows an effective lifetime of atoms around 100 ms. The power of two Raman lasers at the atomic position is 45 mW and 53 mW and the beam waist is 375 $\mu m$ and 150 $\mu m$, respectively. With this configuration, the maximum effective Rabi frequency we reached is about $\Omega_{max}=2\pi\times28.5 $ kHz. During the experiment, the relative phase $\varphi$ between two Raman lasers are also controlled via the phase of radio-freqency that driving the AOM-80MHz. The switch timing and intensity modulation of both Raman lasers are realized by another AOM-110MHz, which is shared by both lasers.

A typical experiment sequence begins with a well prepared ultracold atomic BEC with all atoms populated in the state $|{1}\rangle$. We firstly set the Raman lasers with a large enough two-photon detuning $\delta_0\approx\Omega_{max}$ and then adiabatically turn on the Raman laser while tuning the two-photon detuning $\delta$ according to an adiabatic curve as detailed in the following section. Eventually the Raman laser is kept with target Rabi frequency $\Omega_{eff}(\mathbf{r})$ and $\delta(\mathbf{r})$.
In each experiment sequence, the specified Hamiltonian ${H}_\text{TV}(\mathbf{r})$ is realized and then the spin state of atom is measured by the quantum state tomography for detecting the winding numbers and ploting the spin textures.

\subsection{Adiabatic realization of ${H}_{\text{AL}}(\mathbf{r})$}
\begin{figure}[htbp]
{\includegraphics[height=2.4cm]{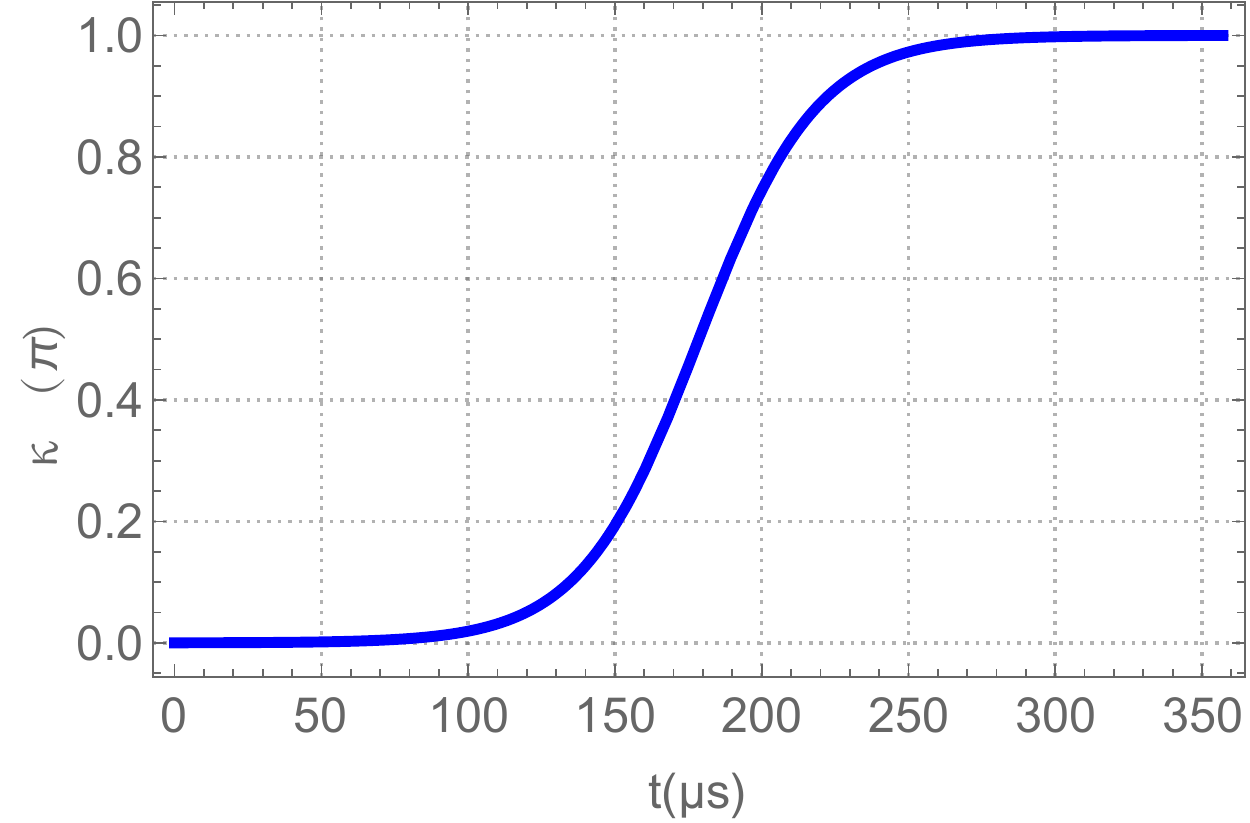}}
{\includegraphics[height=2.4cm]{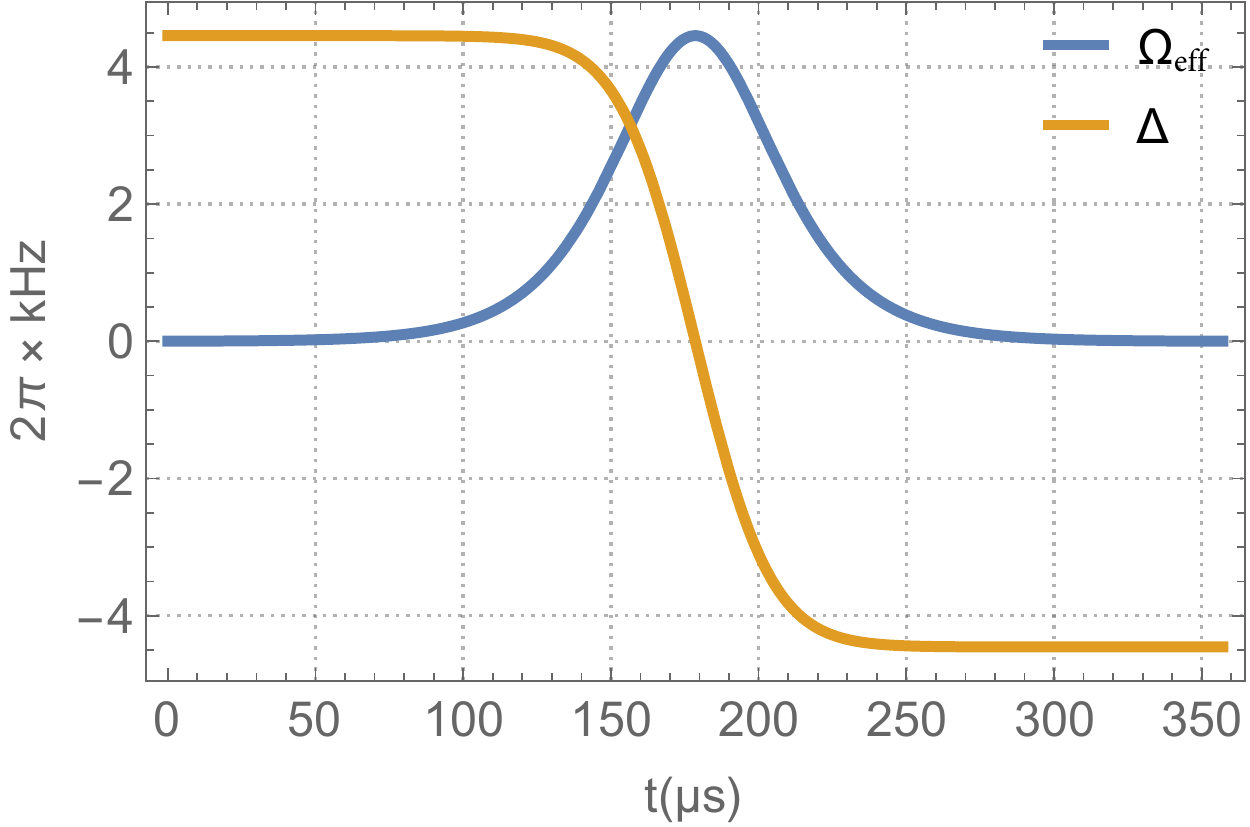}}
\caption{A typical adiabatic passage with final mixing angle $ \kappa $ of $ \pi $. (a) $\kappa$ is changed from 0 to a final value of $ \pi $ following a $\tanh$ curve. (b) An examplary control curve of $\Omega_{eff} $ and $\delta $ in real experiment.}
\label{fig:AdiabaticCurve}
\end{figure}

Here, we show how to adiabatically prepare the atomic quantum state. Under the condition of large single photon detuning $\Delta$, the effective two-level Hamiltonian is shown in Eq.\ref{Hexp}. During the adiabatic process, the instant Hamiltonian would generally take the following form:
\begin{widetext}
\begin{equation}
	{H}_{eff}(t)=\dfrac{\hbar\sqrt{\delta^2(t)+\Omega_{eff}^2(t)}}{2}
	\begin{pmatrix}
		\cos[\kappa(t)]& \sin[\kappa(t)] e^{-i\varphi_0} \\
		\sin[\kappa(t)] e^{i\varphi_0} & -\cos[\kappa(t)]
	\end{pmatrix}
\end{equation}
\end{widetext}
where the mixing angle is defined as $\kappa(t)=\arctan[\Omega_{eff}(t)/\delta(t)]$. The eigen-energy of this effective Hamiltonian can be easily obtained as $E_{\pm}(t)=\pm\hbar/2\sqrt{\delta^2(t)+\Omega_{eff}^2(t)}$, which have much smaller energy difference than $\Delta$ and thus make the near degeneracy approximation always valid.

For each adiabatic state preparation, we start from $ \kappa_0\approx0$ and then increase $\kappa(t)$ slowly to the final value $\kappa_{end}$ in the form of  $\kappa(t)=\kappa_{end} \lbrace\tanh[b (t-T/2)]+1\rbrace/2$ by slowly ramping $\Omega_{eff}(t)$ and $\delta(t)$ following designed control curve. Here $b$ is a parameter determine the ramping speed and eventually the fidelity of prepared quantum state. $T$ defines the time that an adiabatic process last. Each process of state preparation takes $ T =20\pi/\Omega_{max} $, which is 350 $\mu s$ typically. During this adiabatic process, we set the Rabi frequency and two-photon detuning as $\Omega_{eff}(t)=\Omega_{max}\sin\kappa(t)$ and $\delta(t)=\Omega_{max}\cos\kappa(t)$, respectively (See Fig.~\ref{fig:AdiabaticCurve}). Experimentally, $ \Omega_{eff}(t)$ and $ \delta(t)$ are both well controled by amplitude modulation and frequency modulation via AOM-110MHz and AOM-80MHz, respectively. The end value $\kappa_{end}$ and $\varphi_{0end}$ are determined by ${U}_n(\mathbf{r})$ in the following way:
\begin{widetext}
\begin{equation}
\dfrac{\hbar\sqrt{\delta^2_{end}+\Omega^2_{end}}}{2}
	\begin{pmatrix}
		\cos[\kappa_{end}]& \sin[\kappa_{end}] e^{-i\varphi_0} \\
		\sin[\kappa_{end}] e^{i\varphi_0} & -\cos[\kappa_{end}]
	\end{pmatrix}
	= U_{\text{AL}}\sigma_z U^{-1}_{\text{AL}}
\end{equation}
\end{widetext}
where $U_{\text{AL}}=U_n=(U_1)^n$ is determined by the target non-Abelian gauge field at specific position with topological number $n$. Therefore, we first calculate the quantity of $\kappa_{end}$ and $\varphi_{0end}$ at each position $\mathbf{r}$ and then implement the required Hamiltonian following the above adiabatic process.
\section{Measurement of topological number}

\subsection{Measured topological number dependence on spatial grid size and range}
\begin{figure}[htbp]
\includegraphics[width=8cm]{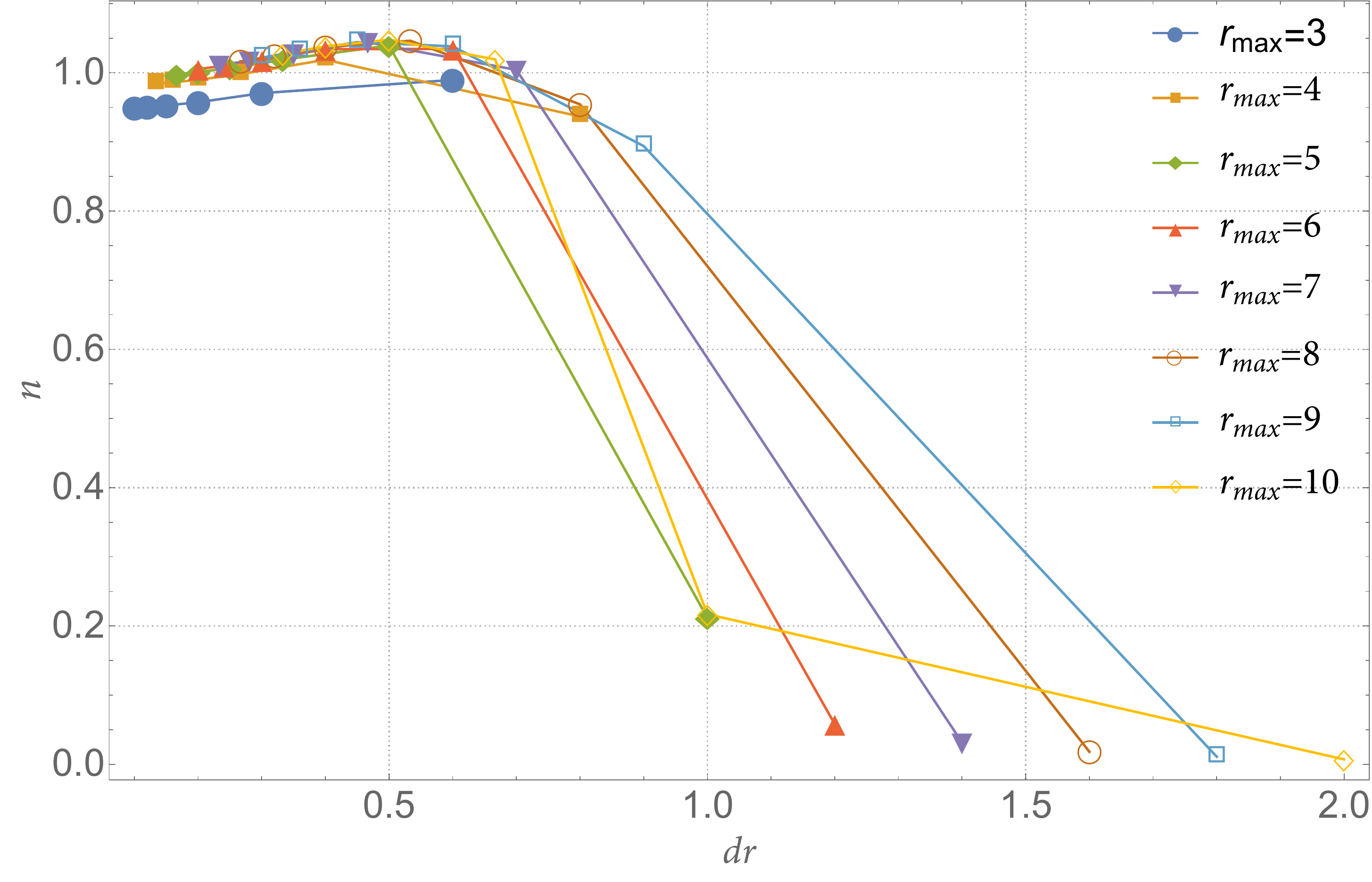}\\
\caption{Numerical results for the Hopf index with different grid sizes $\delta x=\delta y=\delta z=dr$ and grid range $L=[-r_{max},r_{max}]$. As $dr$ becomes smaller and range $r_{max}$ become larger, the  calculated topological number gradually converges to the ideal value 1.}
\label{figgrid_w1}
\end{figure}

Accroding to Eq.\ref{wdef1}, the precision of its discrete version
in experiment is expected to depend on both the grid size and integration range. A dense enough grid and large enough integration range would be important to achieve a correct topological number. We thus try to theoretically determine both proper grid size and grid range before experiment. With smaller grid size $dr$ and larger grid range $r_{max}$, the expected measurable value with the above mentioned detection scheme would give a topological number closer and closer to the ideal value, e.g. $n=1$, as shown in Fig.\ref{figgrid_w1}. Considering that experimentally taking the data of each grid point consumes averagly 300 seconds, in order to finish the topological number measurement with a reasonable error such as less than 10\% in a feasible experiment time, we thus choose the measurement spatial range as $L=[-3\eta,3\eta]$ and $dr=0.1\times(L_{max}-L_{min})$ to measure the topological number of the non-Abelian gauge field $U_1$.

\subsection{Measured topological number dependence on quantum state fidelity}
\begin{figure*}[htbp]
\includegraphics[width=16 cm]{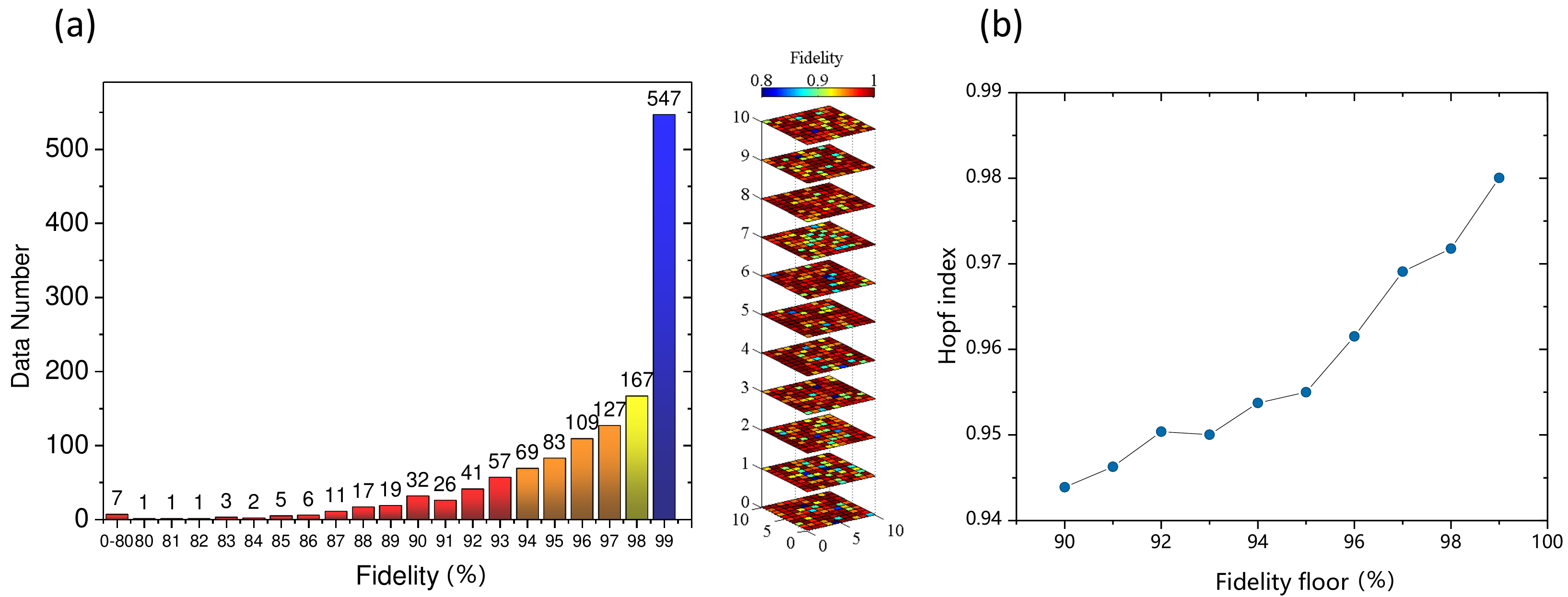}\\
\caption{(a) Measured quantum state fidelity of experimentally prepared quantum state of all the grid sites. Left pannel is the statistics of all  fidelity data while the right pannel is the fidelity dependence on the grid site position. (b) The measured topological number dependence on quantum state fidelity. The fidelity floor means replacing the density matrices with fidelity lower than the fidelity floor by theoretical values. Notice that Hopf index is very close to theoretical value 0.98 when fidelity floor is 99$\%$.}
\label{figFidelity}
\end{figure*}

Another important side to evaluate the experimentally achievable precision of the topological number is the fidelity of experimentally prepared quantum state and also the measured quantum density matrix. Here, the density matrix is reconstructed using the maximal likelihood estimation according to the measured atomic population and thus the expectation value of Bloch vector $\left\langle\sigma_{\{x,y,z\}}\right\rangle$, which is the standard quantum state tomography method. The atomic population in state $|{1}\rangle$ and $|{2}\rangle$ is measured as follows: (1) applying a resonant microwave $\pi$-pulse to transfer all atoms in state $|{2}\rangle$ to state $|{F,m_F}\rangle=|{1,-1}\rangle$; (2) conducting the Stern-Gerlach measurement during the time of flight of atoms; (3) performing the absorption imaging of atoms and count the atom numbers in state $ |{1,-1}\rangle$ and $|{1}\rangle$, which is actually the atomic population in $|{1}\rangle$ and $|{2}\rangle$.

The expectation value of Bloch vector $\left\langle\sigma_z\right\rangle$ can be obtained from the atomic population difference in $|{1}\rangle$ and $|{2}\rangle$. To measure the expectation value of Bloch vectors $\left\langle\sigma_x\right\rangle$ and $\left\langle\sigma_y\right\rangle$, a $\pi/2$ microwave pulse  with phase of 0 or $\pi/2$ is applied to rotate the Bloch vector accordingly before the above population measurement step (1)-(3). As the adiabatic control requiring the tuning of two-photon resonance $\delta$ that acts as an effective magnetic field $B_{z,eff}$ in the Bloch sphere space causing the extra rotation of Bloch sphere, the directly measured Bloch vectors $\left\langle\sigma_{\{x,y,z\}}\right\rangle$ are rotated around the z-axis with an angle of $-\int\delta(t)dt $ before we do the density matrix reconstruction.

The typical fidelity of our quantum state preparation and measurement for the toplogical vacuum of $n=1$ are shown in Fig.\ref{figFidelity}(a), where a statistics and its $\mathbf{r}$ dependence with the above adiabatic procedure is presented. It is clear that the fidelity shows a distribution close to half Gaussian profile while depending weakly on $\mathbf{r}$, which is expected by assuming the random laser mainpulation error. Futhermore, the influence of the infidelity on the measured topological number is presented by replacing the data point below certain fidelity floor by theoretical data and then recalculate the topological number, as shown by Fig.\ref{figFidelity}(b). Here, the topological number obtained using all experimental data is 0.91, while the topological number is
above 0.98 when using all the theoretical data. And by removing the low fidelity date, it is easy to see that higher fidelity would promise a more accurate topological number.

\begin{figure*}[htbp]
\includegraphics[width=15cm]{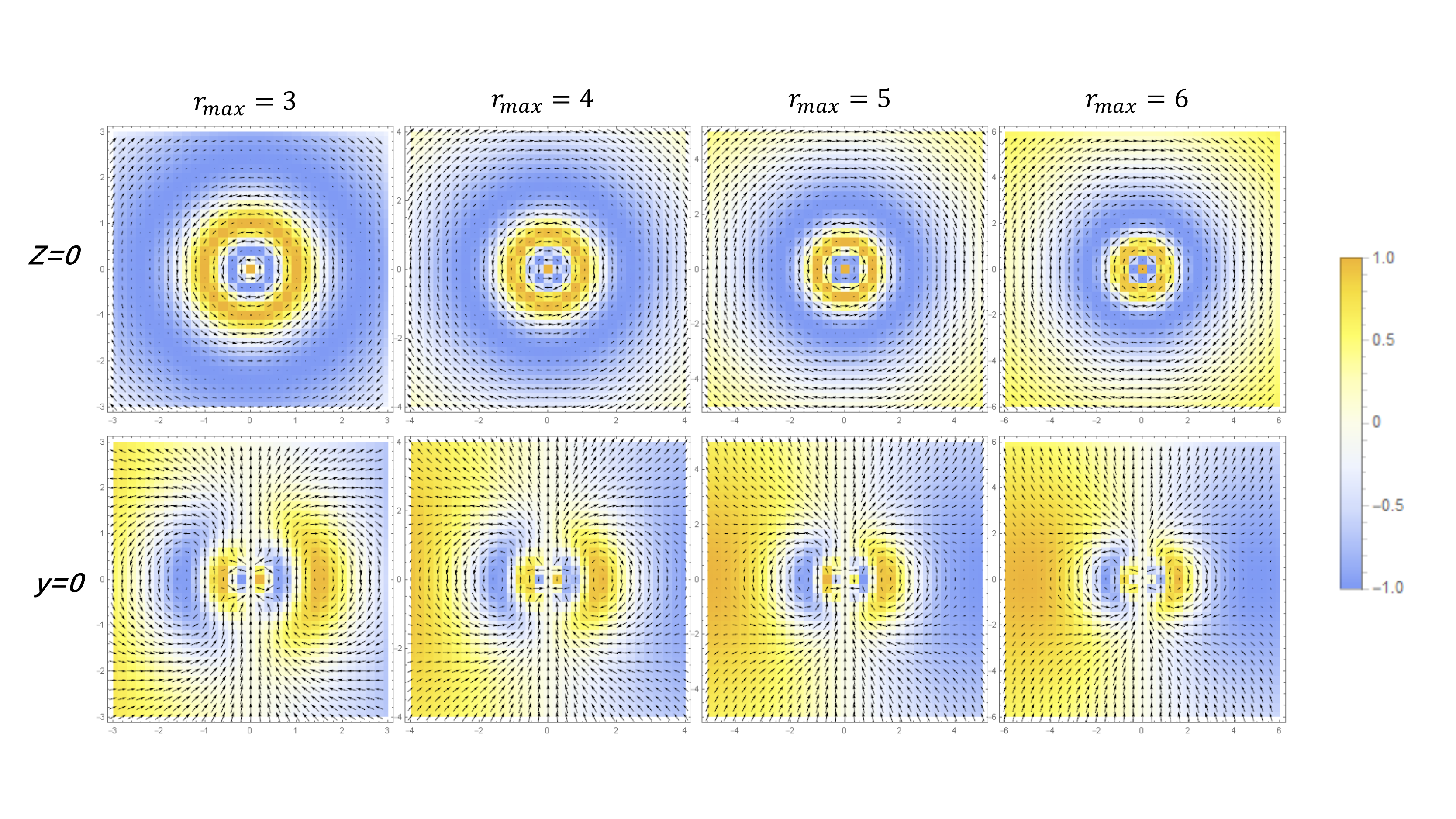}\\
\caption{Spin textures of topological vacuum $n=2$  in  the plane $z=0$ and $y=0$. The grid number $N_{\text{grid}}$ is fixed at $30$ while different grid range $r_{max}$ are compared.}
\label{SpinTextureViaRrange}	
\end{figure*}

\begin{figure*}[htbp]
\includegraphics[width=15cm]{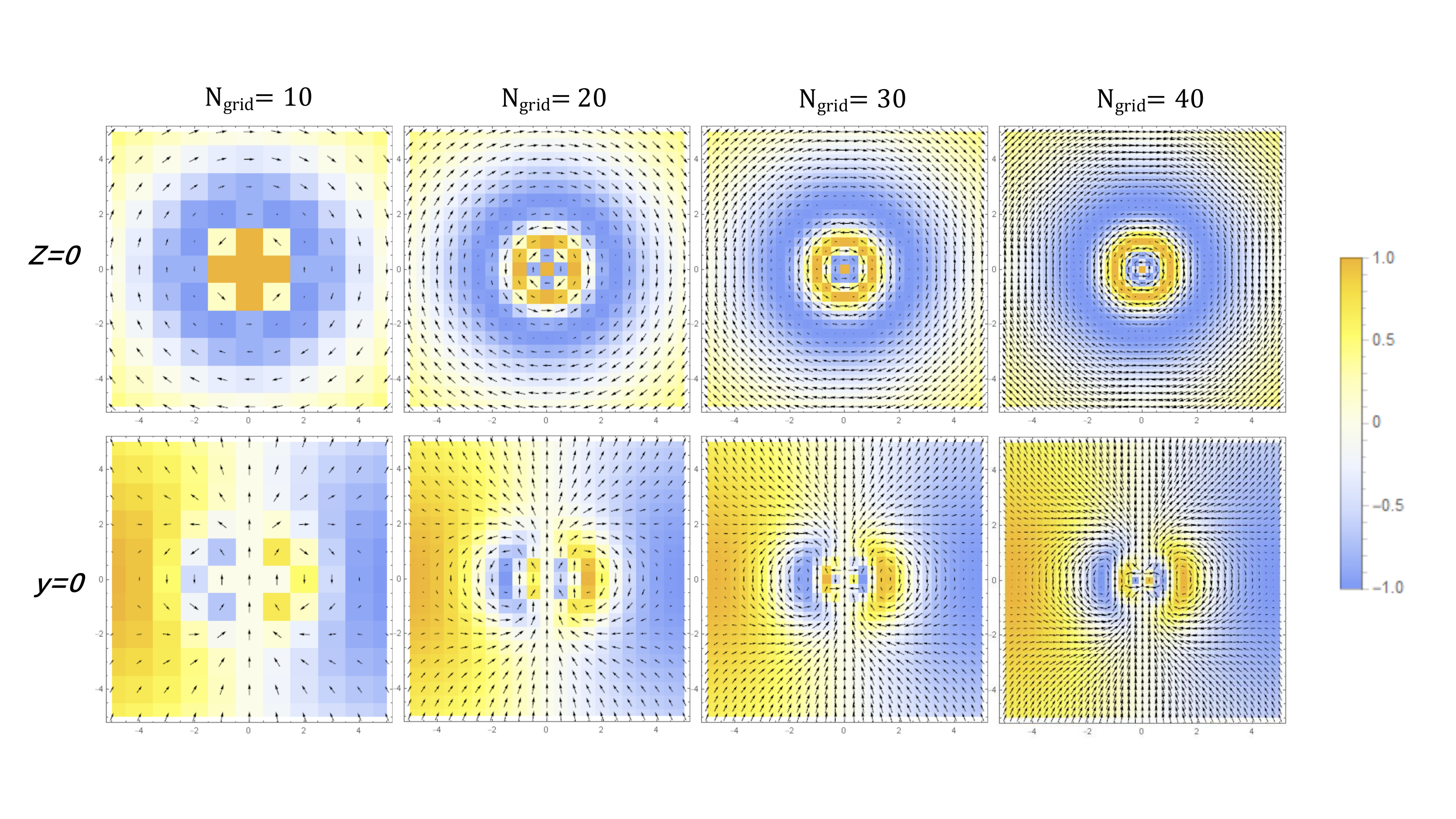}\\
\caption{Spin textures in plane $z=0$ and $y=0$ for the vacuum of  topological number $n=2$. The grid range is fixed at $r_{max}=5$ while different grid numbers $N_{\text{grid}}$ are compared.}
\label{SpinTextureViaGrindN}
\end{figure*}

\section{Measuring spin textures}

Spin texture is a straightforward way to visualize the spatial structure of the Yang-Mills topological vacuum. Theoretically, the winding number can be directly counted from  spin textures. We thus firstly determine the grid density and grid range in a numerical way. As shown in Fig.\ref{SpinTextureViaRrange} and Fig.\ref{SpinTextureViaGrindN}, Bloch vector fields corresponding to the state $|\psi_-(\mathbf{r})\rangle$ are theoretically plotted with different grid size and grid range. As spin textures show much complicated  when topological number $n$ gets larger, we here determine the proper grid size and range according to the topological vacuum with $n=2$. In the main text, the measured spin textures of both $n=1$ and $n=2$ are presented with the same grid size and grid range.

The Bloch vector at specified grid site $\mathbf{r}$ is measured by firstly adiabatically preparing the state and then reconstructed via quantum state tomography. Instead of ploting the Bloch vector field in 3-dimension, we plot its projection on the plane $y=0$ and $z=0$ for the sake of implementing the whole measurement in a reasonable experiment time. Obviously, finer grid resolution and larger grid range would offer clearer spin texture, which, however, increases the data amount required. Eventually, in the experiment, we set the grid range $r_{max}=5$ and grid number $N_{\text{grid}}=30$, which is able to show the topological structure with enough $\mathbf{r}$-resolution.

\begin{figure*}[htbp]
\includegraphics[width=15cm]{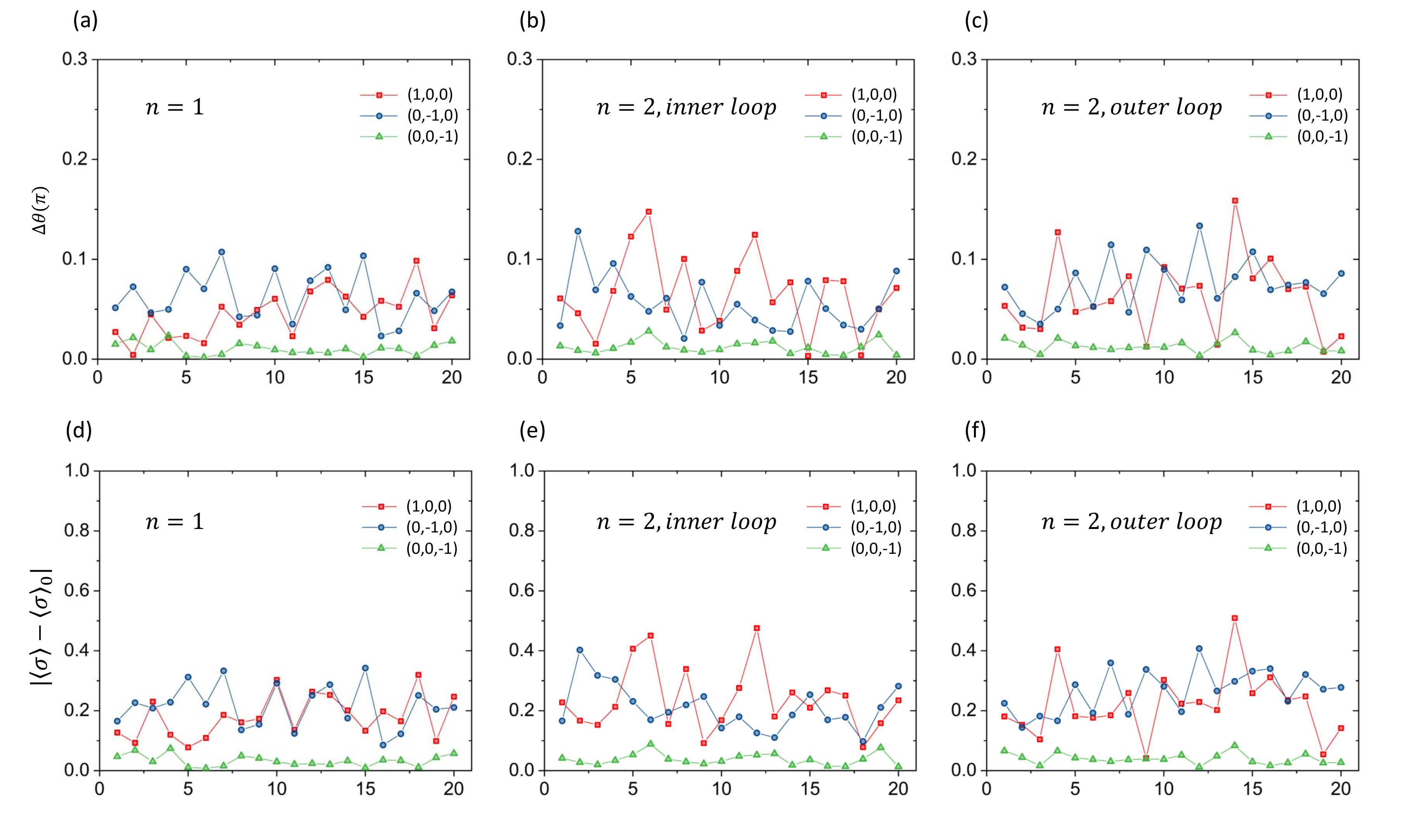}\\
\caption{Deviation between the measured and theoretical spin vector. (a)-(c) show the relative angle $\Delta\theta$ between theoretical and experimental spin orientations on the Hopf link of $n=1$, inner and out loop of $n=2$, repectively. (d)-(f) show the vector distance between theoretical and experimental spin Bloch vectors. Here the horizontal axis number is data point index.}
\label{HopfLinkErr}
\end{figure*}

\section{Measuring the Hopf links}

Hopf links are also another direct way to visualize the topology of synthetic vacua. For a topological vacuum with winding number $n$ in the real case, the position where spin vector points to the same direction such as $+\vec{z}=\{0,0,1\} $ would form $n$ cocentric loops. Moreover, loops belong to different spin orentations would interwind with each other. Therefore, Hopf-links offer us another direct way to evaluate the topological number of a Yang-Mills vacuum. However, in order to find these Hopf-links, we have to in principle measure the Bloch vectors in 3 dimensional space with extremely high grid resolution and grid range.

In experiment, the spin orientation $\langle {\mathbf{\sigma}}\rangle$ are always measured with finite error. Considering a small $\epsilon$-neigborhood of a specific orientation (e.g., $\langle {\mathbf{\sigma}}\rangle_1$), we may define:
\begin{equation}
N_{\epsilon}(\langle {\mathbf{\sigma}}\rangle_1)=\{\langle {\mathbf{\sigma}}\rangle:|\langle {\mathbf{\sigma}}\rangle-\langle {\mathbf{\sigma}}\rangle_{0}|<\epsilon \},
\end{equation}
where
$|\langle {\mathbf{\sigma}}\rangle-\langle {\mathbf{\sigma}}\rangle_{0}|=
\sqrt{( \langle {\mathbf{\sigma}}\rangle_x- \langle {\mathbf{\sigma}}\rangle_{0x})^2
+(\langle {\mathbf{\sigma}}\rangle_y- \langle {\mathbf{\sigma}}\rangle_{0y})^2
+(\langle {\mathbf{\sigma}}\rangle_z- \langle {\mathbf{\sigma}}\rangle_{0z})^2}$
represents the distance between $\langle {\mathbf{\sigma}}\rangle$ and $\langle {\mathbf{\sigma}}\rangle_0$. For example, with $\epsilon=0.01$, a grid number $N_{\text{grid}}=80$ would only promise us to find $N_{\epsilon}=20$ points on the Hopf-link. The amount of experimental data required to measure in 3 dimension reach up to $(N_{\text{grid}}+1)^3=531441$, which is hardly finished experimentally in a reasonable time. Therefore, we instead theoretically find the position $\mathbf{r}$ on Hopf-links and then experimentally measure the spin vectors $\langle {\mathbf{\sigma}}\rangle(\mathbf{r})$ to verify the existence of Hopf-link.

We eventually measure $N_{\epsilon}=20$ spin vectors $\langle {\mathbf{\sigma}}\rangle(\mathbf{r})$ on each Hopf-link that evenly locate on the loop for both topological number $n=1$ and $n=2$.  Beside the above defined distance $|\langle {\mathbf{\sigma}}\rangle-\langle {\mathbf{\sigma}}\rangle_{0}|$, the relative angel $\Delta\theta$ spanned between the experimental $\langle {\mathbf{\sigma}}\rangle$ and theoretical $\langle {\mathbf{\sigma}}\rangle_{0}$ are also calcualted. Corresponding to the Hopf-link presented in the main text, the relative angle $\Delta\theta$ and vector distance $|\langle {\mathbf{\sigma}}\rangle-\langle {\mathbf{\sigma}}\rangle_{0}|$ of each experimental data are shown in Fig.\ref{HopfLinkErr}.  Averagely, the relative angle we achieve is $\Delta\theta\approx0.05\pi$ while the $\overline{|\langle {\mathbf{\sigma}}\rangle-\langle {\mathbf{\sigma}}\rangle_{0}|}$ is shown in the Table \ref{linkErrTbl}. The deviation $\overline{|\langle {\mathbf{\sigma}}\rangle-\langle {\mathbf{\sigma}}\rangle_{0}|}$ for the loop $(0,0,1)$ is much smaller than that  of other two loops since spin vector $\langle {\mathbf{\sigma}}\rangle(\mathbf{r})$ is exactly the initial state for  this loop. In other words, the required eigenstate of $H_{TV}$ is just the initial state and thus the errors from adiabatically driving $H_{TV}$ vanish for the loop $(0,0,1)$.
\begin{table}
\centering
\caption{\label{linkErrTbl}spin vector deviation $\overline{|\langle {\mathbf{\sigma}}\rangle-\langle {\mathbf{\sigma}}\rangle_{0}}$}
\begin{tabular}{|l|l|l|l|}
\hline
Hopf-link loop & (-1,0,0) & (0,1,0) & (0,0,1)  \\
\hline
$n=1$                    & 0.18       & 0.22      & 0.03       \\
\hline
$n=2$ inner loop         & 0.24       & 0.21      & 0.04       \\
\hline
$n=2$ outer loop         & 0.22       & 0.26      & 0.04       \\
\hline
\end{tabular}
\end{table}

\end{appendix}


\end{document}